\documentclass{aa}

\usepackage{graphicx}
\usepackage{txfonts}
\usepackage{xcolor}
\usepackage{amsmath}
\usepackage{orcidlink}
\usepackage[normalem]{ulem}
\newcommand{\kms}{$\mathrm{km\,s^{-1}}$}
\newcommand{\vsini} {$v\sin\,i$}
\DeclareRobustCommand{\Eqref}[1]{Equation~\ref{#1}}
\DeclareRobustCommand{\Figref}[1]{Figure~\ref{#1}}
\DeclareRobustCommand{\Tabref}[1]{Table~\ref{#1}}
\DeclareRobustCommand{\Secref}[1]{Section~\ref{#1}}
\DeclareRobustCommand{\Appref}[1]{Appendix~\ref{#1}}

\usepackage{showyourwork}
\usepackage{hyperref}

\begin{document}

\graphicspath{{./figures/}}

\title{Tracing the evolution of short-period binaries with super-synchronous fast rotators}

\author{N. Britavskiy\inst{1,2}\orcidlink{0000-0003-3996-0175},
  M. Renzo\inst{3,4}\orcidlink{0000-0002-6718-9472},
  Y.~Naz\'e\inst{1}\thanks{FNRS Senior Research Associate}\orcidlink{0000-0003-4071-9346},
  G.~Rauw\inst{1}\orcidlink{0000-0003-4715-9871},
  P.~Vynatheya\inst{5}\orcidlink{0000-0003-3736-2059}
}

\institute{Universit\'e de Li\`ege, Quartier Agora (B5c, Institut d’Astrophysique et de Geophysique), All\'ee du 6 Ao\"ut 19c, B-4000 Sart Tilman, Li\`ege, Belgium \and Royal Observatory of Belgium, Avenue Circulaire/Ringlaan 3, B-1180 Brussels, Belgium
\and Steward Observatory, University of Arizona, 933 N. Cherry Avenue, Tucson, AZ 85721, USA \and
  Center for Computational Astrophysics, Flatiron Institute, New York, NY 10010, USA \and
  Max-Planck-Institut für Astrophysik, Karl-Schwarzschild-Strasse 1, 85748 Garching bei München, Germany
}
\offprints{mbritavskiy@uliege.be}

\date{Submitted 06 November 2023 / Accepted 13 January 2024}

\titlerunning{Fast-rotating massive stars in a short-period binaries}

\authorrunning{N. Britavskiy et al.}


\abstract
{The initial distribution of rotational velocities of stars is still
  poorly known, and how the stellar spin evolves from birth to the
  various end points of stellar evolution is an actively debated
  topic. Binary interactions are often invoked to explain the
  existence of extremely fast-rotating stars (\vsini~
  $\gtrsim 200$\,\kms). The primary mechanisms through which binaries
  can spin up stars are tidal interactions, mass transfer, and
  possibly mergers. However, fast rotation could also be primordial,
  that is, a result of the star formation process.
  To evaluate these scenarios, we investigated in detail the
  evolution of three known fast-rotating stars in short-period spectroscopic and eclipsing
  binaries, namely HD\,25631, HD\,191495, and HD\,46485, with
  primaries of masses of 7, 15, and 24 $M_{\odot}$, respectively, with
  companions of $\sim1\,M_\odot$ and orbital periods of less than 7 days. These systems belong to a
  recently identified class of binaries with extreme mass ratios, whose evolutionary origin is still poorly understood.}
{We evaluated in detail three scenarios that could explain the fast
    rotation observed in these binaries: it could be primordial, a
    product of mass transfer, or the result of a merger within an
    originally triple system. We also discuss the future evolution of
    these systems to shed light on the impact of fast rotation on
    binary products.}
  {We computed grids of single and binary MESA models varying tidal
      forces and initial binary architectures to investigate the
      evolution and reproduce observational properties of these
      systems. When considering the triple scenario, we determined the
      region of parameter space compatible with the observed binaries
      and used a publicly available machine-learning model to determine
      the dynamical stability of the triple system.}
{We find that, because of the extreme mass-ratio between binary components,
     tides have a limited impact, regardless of the prescription used, and that
    the observed short orbital periods are at odds with post-mass-transfer
    scenarios.
   We also find that the overwhelming majority of triple systems compatible with the observed binaries are dynamically unstable and would be disrupted within years of formation, forcing a hypothetical merger to happen so close to a zero-age main-sequence that it could be considered part of the star formation process.}
 {The most likely scenario to form such young, rapidly rotating,
    and short-period binaries is primordial rotation, implying that the
    observed binaries are pre-interaction ones. Our simulations further indicate
  that such systems will subsequently go through a common envelope and
  likely merge. These binaries show that the initial spin distribution of massive
  stars can have a wide range of rotational velocities.}
{}


\maketitle
\section{Introduction}

Stellar rotation is a ubiquitous though as of yet poorly understood
phenomenon, especially for massive stars \cite[e.g.,][]{Langer_2012}.
Rotation is crucial for the nucleosynthesis of certain elements
\citep[e.g., ``s process''][]{limongi:18}, for powering transients
such as long gamma-ray bursts in the ``collapsar'' scenario
\citep[e.g.,][]{macfadyen:99}, and to understand the spins of neutron
stars and black holes \citep[e.g.,][]{callister:21, GWTC3}.
Understanding the rotational evolution of stellar surfaces and
interiors is thus important for galactic evolution, time domain astrophysics, and
gravitational-wave observations. Before investigating the origin of stellar
rotation, we shall focus on how the rotation is
studied and what affects it.

\paragraph{Observational constraints on stellar rotation.}
For O- and B-type stars, spectroscopic observations typically provide only the projected
  equatorial rotational velocity \vsini, that is, the equatorial rotational
  velocity $v$ times the sine of the inclination angle to the line of
  sight $i$, which is usually unknown.
Typically, \vsini~ is measured through the Doppler broadening of spectral
lines, where the rotational component needs to be carefully separated
from other line-profile broadening components (e.g., macroturbulence; \citealt{Simon-Diaz2014}).

While investigating various stellar populations, common features in
the distribution of \vsini~ are found for O-type stars
\citep{Conti_1977, vfts_2013_otype, vfts_2015_otype}, B-type
stars \citep{Wolff_1982, Dufton_2013}, and A-type stars \citep{Abt_1973}.
The distribution of \vsini~ does indeed seem to peak near 100~\kms~ with
a tail of so-called fast-rotating stars with \vsini\,$>$\,200~\kms.

Recent studies describing large surveys (including stellar
populations in the Magellanic Clouds) have confirmed this overall
morphology of the \vsini~distribution of early-type massive stars
\citep[e.g.,][]{Dufton_2013,vfts_2013_otype,Holgado_2022}. Notably,
the spin distribution appears independent of the environment
\citep[e.g., Galactic field vs. young stellar
clusters;][]{Wolff_2007,Penny_2009, Huang_2010}. However, field stars
are older than stars in young open clusters, on average, and since
rotation slows down as the star evolves because of stellar winds and the evolutionary expansion of the envelope, stars in a young open cluster appear to rotate
systematically faster than stars in the field.
Interestingly, in more metal-poor environments, there is a trend
toward increasing mean rotational velocity in OB stars while keeping
the same overall nonuniform distribution \citep{Ramachandran_2019}. This is caused by the
  weakening of stellar winds at low metallicity
\citep[e.g.,][]{Vink_2001,mokiem:07}: the initial angular momentum of
the star is preserved longer. These results tentatively imply that the
observed \vsini~ distribution exists in environments of differing metallicity
and age.

The shape of the surface rotational distribution of massive
  stars is determined by various poorly understood processes, such as the unknown pre-main-sequence (pre-MS)
  initial distribution and the influence of various physical processes
  active as the stars evolve and possibly interact with commonly
  occurring \citep[e.g.,][]{Sana_2012, moe_2017, Offner_2022} stellar
  companions \citep[e.g.,][]{packet:81, deMink_2013, vfts_2015_otype}.
This nonuniform (also referred as bimodal) rotational distribution is causing the MS split which is observed on the color magnitude diagrams of the young star clusters \citep[see, for example,][]{Kamann_2023,Wang_2023}.

\paragraph{What affects the stellar rotation?}
The main factors that can affect stellar rotation are as follows: (\emph{i}) the
angular momentum budget left behind by star formation processes;
(\emph{ii}) the acting stellar physics processes throughout the
  evolution elapsed, for example, magnetic braking, mass loss, changes in internal structure; and
(\emph{iii}) binary interactions, including stellar mergers.

The initial spin of a star is the direct product of star formation
processes, that happen embedded in opaque clouds. Different physical
aspects have been suggested to play a major role, such as the magnetic
field or the presence of a circumstellar disk through the pre-MS stage
\citep[e.g.,][]{Rosen_2012}. In fact, these two factors are
proposed to be dominant in producing the nonuniform distribution
of rotation among the pre-MS low- and intermediate-mass stars
\citep{Bastian_2020}. This initial rotation distribution has not been
studied in the pre-MS massive star domain; however, based on the similarity of the observed \vsini~ distribution across spectral types, we can expect that a similar
 non-uniform initial distribution could also be present there.

After formation, the spin evolution depends on highly debated plasma physics (e.g., shears and magnetic
instabilities; see e.g., \citealt{Tayler_1973, spruit:99, spruit:02,Maeder_2000, fuller:19, denhartogh:20,ji:23}), uncertain wind mass
and angular momentum loss rates \citep[e.g.,][]{smith:14, renzo:17},
and the evolution of the stellar moment of inertia \citep[e.g.,][]{langer:98,zhao:20}.
Each of these evolutionary processes can modify the shape of the observed distribution of surface rotation \citep{Maeder_2000}.

Uncertainties affecting these stellar processes limit our
understanding of the angular momentum budget of a star in late evolutionary phases.
At core-collapse, hydrodynamical instabilities can redistribute this angular momentum between the newly formed compact object and the ejecta \citep[e.g.,][]{Kazeroni_2017}, further complicating the
connection between observed compact object spins and their parent stars' rotation.

In addition, the vast majority of massive stars exist in multiple systems \citep[e.g.,][]{Mason_2009, Duchene_2013, Sana_2013, Kobulnicky_2014, Dunstall_2015, Almeida_2017, Offner_2022}.
In such systems, interactions among stars are expected to occur in 70 \% of cases during the MS stage evolutionary stage \citep{Sana_2012}.
Such binary interactions may also play a dominant role in shaping the observed
distribution of the surface rotation rate, especially for the tail of fast rotators because of tides, mass transfer, or mergers. This idea has been proposed
  multiple times, both from a theoretical perspective
  \citep{packet:81, Pols_1994, deMink_2013, vinciguerra:20} and observational studies \citep{blaauw:93,  vfts_2015_otype, Cazorla2017b, Britavskiy_2023}.
One example is the famous category of Be stars, which are fast-rotating stars with decretion disks \citep{Fremat_2005, Rivinius_2013}.
They are not known to be paired with MS companions \citep{Bodensteiner_2020}, but instead, many of them are orbited by a stripped star \citep{Wang_2018,Wang_2021} or a compact companion \citep[e.g.,][]{Reig_2011,Dodd_2023}.

However, even if binarity causes the fast-rotating tail of the distribution, this still leaves the nonnegligible part of the distribution unexplained. For example, \citet{vfts_2015_otype} suggest that
the initial rotation distribution of massive stars is the same for
apparently single and single-lined spectroscopic binary (SB1) stars.
This gives us a hint that some of the rapid rotators can be effectively single or still be in the pre-interacting binary regime.
Conversely, some of the young spectroscopic binaries are still in the pre-interacting stage and the rotation of its components has not yet been affected by binary interaction. A few examples have been proposed in the past \citep{Johnston2021,Stassun_2021, Naze_2023_rot}.
To distinguish what exactly causes the fast rotation, we need to investigate different binary configurations and their evolution from the rotation point of view 
in detail.

To summarize, stellar rotation could indirectly indicate whether a binary interaction has happened in a given system at some point in its evolution.
To consider rotation as a tracer of a binary interaction, it is however important to understand which factors affect it.
Answering these questions will help us to understand the overall evolution of stellar spin, from stellar birth to the latest various end points of stellar evolution.

\paragraph{Main motivation of the present work.}
Here, we study known Galactic fast-rotating OB stars that have been identified
  in single-lined spectroscopic binaries, specifically HD\,25631,
  HD\,191495, and HD\,46485
  \citep[SB1;][see also \Tabref{table:sb1_fast}]{Britavskiy_2023,Naze_2023_rot}. Taking into account
  their high surface rotational velocities (\vsini\,$>$\,200~\kms), short orbital periods ($P$ $<$ 7 days), and very unequal
  mass ratios, we aim to understand what causes such a high rotation, and to determine what the effect is (if any)
of past and future binary interactions. Their observed \vsini~ is indeed significantly larger than the tidal synchronization velocities.
These systems are not exceptional: similar short-period
binaries with an extreme mass ratio are known \citep[][]{Moe_2015,Jerzykiewicz_2021}. Thus, by evaluating different processes that
could result in a high rotation for the investigated targets, we can
emphasize what  the more probable scenario is to form the whole class of such binaries.
The main goal of the current work is to evaluate different evolutionary scenarios that could result in these peculiar binary configurations and to predict how they will further evolve.

Numerical calculations of the internal evolution of both stars in an
interacting binary accounting for the exchange of mass and angular
momentum \citep[e.g.,][]{wang:20, Badry_2021,Renzo_2021, pauli:22,
  renzo:23} allowed us to investigate the detailed evolution of
rotational velocities in such systems. We followed the evolution of these targets up to late stages, in order to understand whether they will be the sources of stripped stars, stellar mergers, and/or X-ray binaries.

The paper is organized as follows. In \Secref{sec:sample}, we introduce the sample of fast rotators and the methodology we followed; from \Secref{sec:pre-interaction} to \Secref{sec:triple_scenario}, we evaluate three different scenarios of binary interaction that could occur in a given system.
The primordial rotation scenario (pre-interacting regime) is presented
in \Secref{sec:pre-interaction}. We evaluate how long a high
rotational velocity can remain, assuming that these systems were formed
with an initial rotation close to the current \vsini. In addition, we highlight what the effect of tides is on the rotation and which further binary interaction will affect such systems.
In \Secref{sec:post-interaction} we evaluate if the present orbital configuration of the three systems could form through a binary interaction. We assume that the accretor has gained its high rotational velocity via stable mass transfer from the donor.
In \Secref{sec:triple_scenario} we present our discussion regarding a merger in a triple system. A merging process could result in a fast-rotating star; however, dynamic stability parameters that are able to reach the observed binary configurations need to be investigated.
\Secref{sec:discussions} presents a general discussion, while \Secref{sec:conclusions} recalls the main conclusions of this paper.

\section{Sample of stars and the adopted methodology}
\label{sec:sample}

Recently, three peculiar short-period binaries with a fast rotator
(\vsini\,$>$\,200~\kms) as primary component \citep{Naze_2023_rot}
were detected and analyzed in detail. These three targets (HD\,25631,
HD\,191495 and HD\,46485) have common morphology of light-, and
radial-velocity curves, namely two narrow eclipses and a strong
reflection effect through one orbital cycle indicating that the
secondary component has a lower surface temperature than its primary
and is thus \emph{not} a compact object. The orbital and
fundamental physical parameters of these targets have been studied in
detail in \citet{Naze_2023_rot} and are summarized in \Tabref{table:sb1_fast}. Notably, these three SB1 systems
share short orbital periods (3.2 to 6.9 days) and rather extreme
mass ratios ($q=M_2/M_1=M_\mathrm{unseen}/M_\mathrm{seen}$ from 0.035 to 0.15).

\begin{figure}[!ht]
  \centering
  \script{hr_plot_paper2023_fin.py}
  \includegraphics[width=0.50\textwidth]{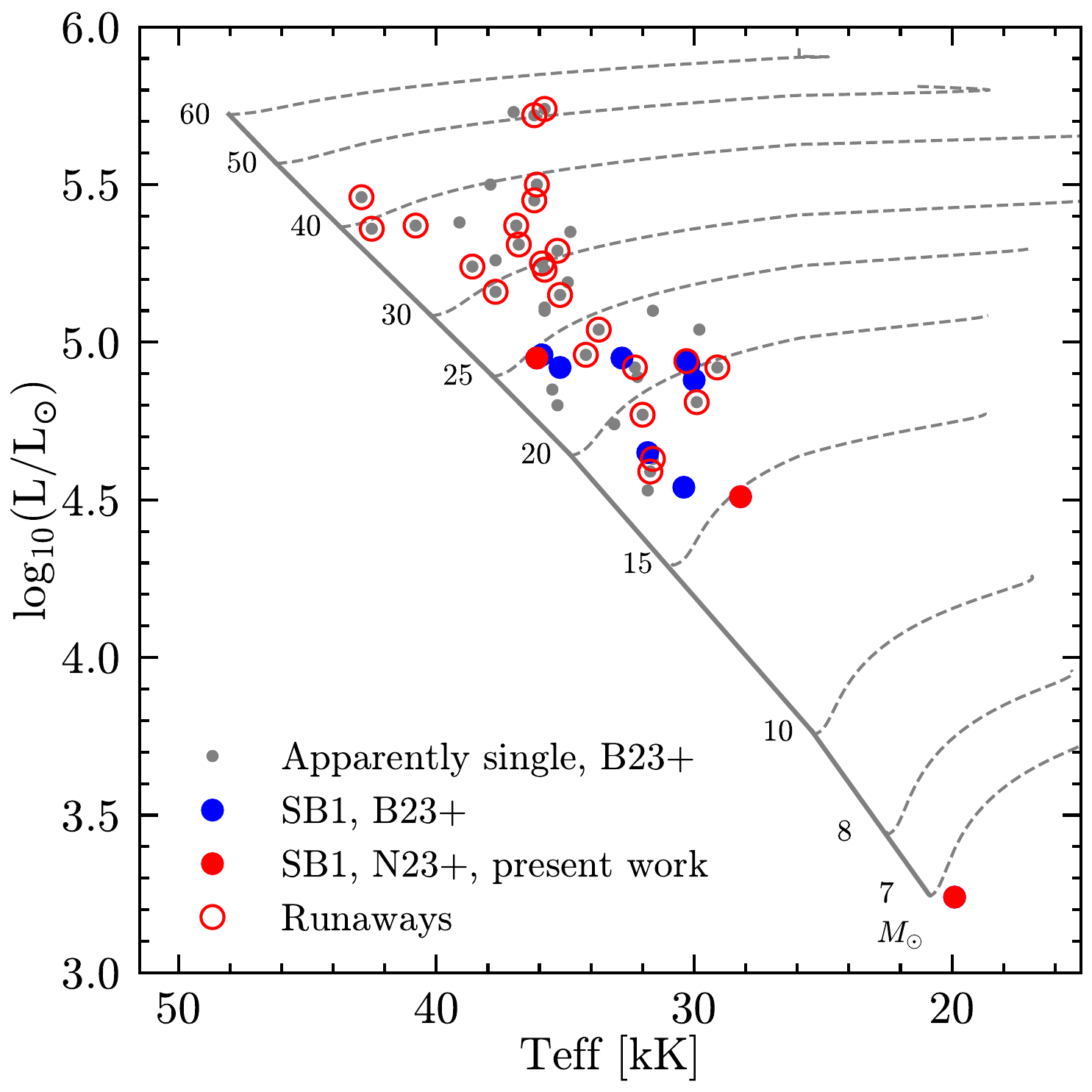}
  \caption{Location in the HR diagram of the Galactic OB fast rotators studied in \citet{Britavskiy_2023} and \citet{Naze_2023_rot}. Apparently single, binary, and runaway fast rotators are marked by specific symbols. Nonrotating evolutionary tracks represent the main-sequence stage of a single-star evolution at solar metallicity for different initial masses.}
  \label{fig:hr_plot_paper2023}
\end{figure}

The position of primaries of the selected systems on the Hertzsprung--Russell (HR)
diagram is presented in \Figref{fig:hr_plot_paper2023} together
with a sample of 50 Galactic fast-rotating O stars from \citet{Britavskiy_2023} for reference.
In addition, we calculated the evolutionary tracks that represent the main-sequence evolutionary stage of a nonrotating single star at solar metallicity (Z = 0.02).
For each track, the zero-age main-sequence (ZAMS) was defined at the moment when total nuclear luminosity from all burning becomes equal to the surface luminosity within the absolute difference of $10^{-4}$.
The end of MS is defined as the central mass fraction of hydrogen decreasing below $X(^{1}\mathrm{H})<0.01$.

The ages for these systems were
derived by fitting the BONNSAI evolutionary tracks \citep{Bonnsai_2014} at Galactic
metallicity to the derived effective temperatures and luminosities of
the primaries. These are in agreement with the age estimates of the
parent clusters for two systems belonging to clusters, namely
HD\,46485 in NGC 2244 and HD\,191495 in NGC 6871 \citep[see for details][]{Naze_2023_rot}.

The existence of these systems raises questions regarding
their origin and evolutionary paths, which can shed light on the initial
\vsini~distribution of OB stars. However, first, we need to understand what is the origin of the super-synchronous rotation of the primaries in these systems.

\begin{table}[!htbp]
{\small
\caption{Basic physical and orbital parameters of three
  fast-rotating SB1 systems that have been investigated by
  \citet{Naze_2023_rot}. }
\label{table:sb1_fast}
\begin{tabular}{l c c c}
\hline
\hline
Target                  & HD\,25631          &  HD\,191495      & HD\,46485 \\
\hline
SpC (primary)             &  B3V             &  B0V             &  O7\,V((f))nvar?  \\
log($L_{1}/L_{\odot}$)    &  3.24 $\pm$ 0.02 & 4.51 $\pm$ 0.04  & 4.95 $\pm$ 0.03 \\
T$_{eff,1}$ [kK]          & 19.9 $\pm$ 0.5$^{\ast}$  & 28.2 $\pm$ 1.3   & 36.1 $\pm$ 0.7  \\
\vsini~ [\kms]            &  221$\pm$ 10 &  201$\pm$ 10  & 334$\pm$ 16  \\
\hline
age  [Myr]       &  9.6$\pm$9.4  &  7.6$\pm$1.8  & 2.5$\pm$1.1   \\
$M_\mathrm{1}$   [$M_{\odot}$]  &  7 -- 8  &  15    & 24  \\
$M_\mathrm{2}$   [$M_{\odot}$] &  0.8 -- 1.2 & 1.2 -- 1.8 &  0.8 -- 1.1 \\
$P$  [d]  &  5.2  & 3.6   &  6.9 \\
$e$       & 0   &  0  &  0.033 \\
$i$ [$\deg$]  &  79 -- 80  &  67 -- 69  &  73 -- 76 \\
$q$ & 0.12 -- 0.15 & 0.08 -- 0.12 & 0.035 -- 0.045 \\
\hline
\end{tabular}
Notes. $^{\ast}$ - the uncertainty in T$_{eff}$ taken as average uncertainty of the method presented in \citet{Braganca_2012}.
}
\end{table}

To model the different evolutionary scenarios we used the MESA
\citep[Modules for Experiments in Stellar Astrophysics, version
15140][]{Paxton2011, Paxton2013, Paxton2015, Paxton2018, Paxton2019,
  Jermyn2023}. The adopted input parameters and output results for our
models are available on GitHub\footnote{\url{https://github.com/NikolayBritavskiyAstro/fast_rotating_binaries}} and Zenodo\footnote{\url{https://zenodo.org/records/10479754}}. As a
starting point we took the setup from \citet{Renzo_2021} based
  on reproducing the Galactic O-type fast rotator $\zeta$ Ophiuchi.

All our models have a solar metallicity Z = 0.02, considering that the observed sample consists of
Galactic stars. In the subsequent sections, we describe the main
physical ingredients relevant to the specific scenario we consider,
\Appref{sec:MESA_setup} describes the full list of physical and
numerical assumptions. Following the "best practice" tips of working with MESA, we repeated calculations by performing resolution tests (see \Appref{sec:resolution_test} for details).

\section{Primordial rotation -- Pre-interaction scenario}
\label{sec:pre-interaction}

In this section, we assume that the primaries of HD\,25631, HD\,191495, and HD\,46485 were
born with a high rotation and we describe their subsequent
evolution.
Our main question is how long such a fast initial rotation can remain - that is, could such initial rotation be detected in practice -  and what is the effect of tides in such short-period systems?

\subsection{Method}

As we are investigating the surface rotational properties of massive stars, it is very important to take into account the main factors that affect it. In the case of single stars, we need to consider assumptions regarding the
mass loss through stellar winds, core overshooting, and expansion of the envelope while the star is evolving. For the consideration of mass loss, we use the \texttt{"Dutch"}
wind scheme that combines \citet{Vink_2001} mass loss prescription,
suitable for effective temperatures $\geq$ 11 000 K and mass fraction
of hydrogen $X(^{1}\mathrm{H})$ $>$ 0.4, with \citet{Jager_1988}
mass loss rates for temperatures lower than 10 000 K. In between the
mentioned temperatures, the \texttt{"Dutch"} scheme is linearly
interpolating hot and cold wind prescriptions.
To test robustness of our conclusions against uncertainties in wind mass loss rates \citep{renzo:17}, we also compute models with the "flat" wind mass loss rate from \citet{Bjorklund_2023}.
 Regarding extended convective mixing, we use the exponential core overshooting based on
\citet{Herwig_2000} with free parameters chosen following
\cite{claret:18} work which reproduces the width of the main-sequence in 30
Doradus predicted by \citet{Brott_2011}. These adopted parameters are essential for a proper definition of stellar
structure (e.g., mass of the core, mass loss, etc) taking into account
that they will affect the rotational properties of the stars under
investigation.
While investigating the evolution of surface rotational velocity, it is important to be aware of which prescription of angular momentum (AM) transport has been adopted, in all our models we assume a Spruit - Tayler dynamo mechanism \citep{spruit:02}.
To test how it affects the surface rotation we ran several models without it and in addition, we ran the prescription of \citet{fuller:19} AM transport mechanism that is more efficient with respect to a Spruit - Tayler dynamo.

\subsection{Results and interpretation (effect of mass loss)}
\label{sec:res_primordial}

\begin{figure}[!ht]
  \centering
  \script{plot_save_mesa_comparison_wind_fin.py}
  \includegraphics[width=0.50\textwidth]{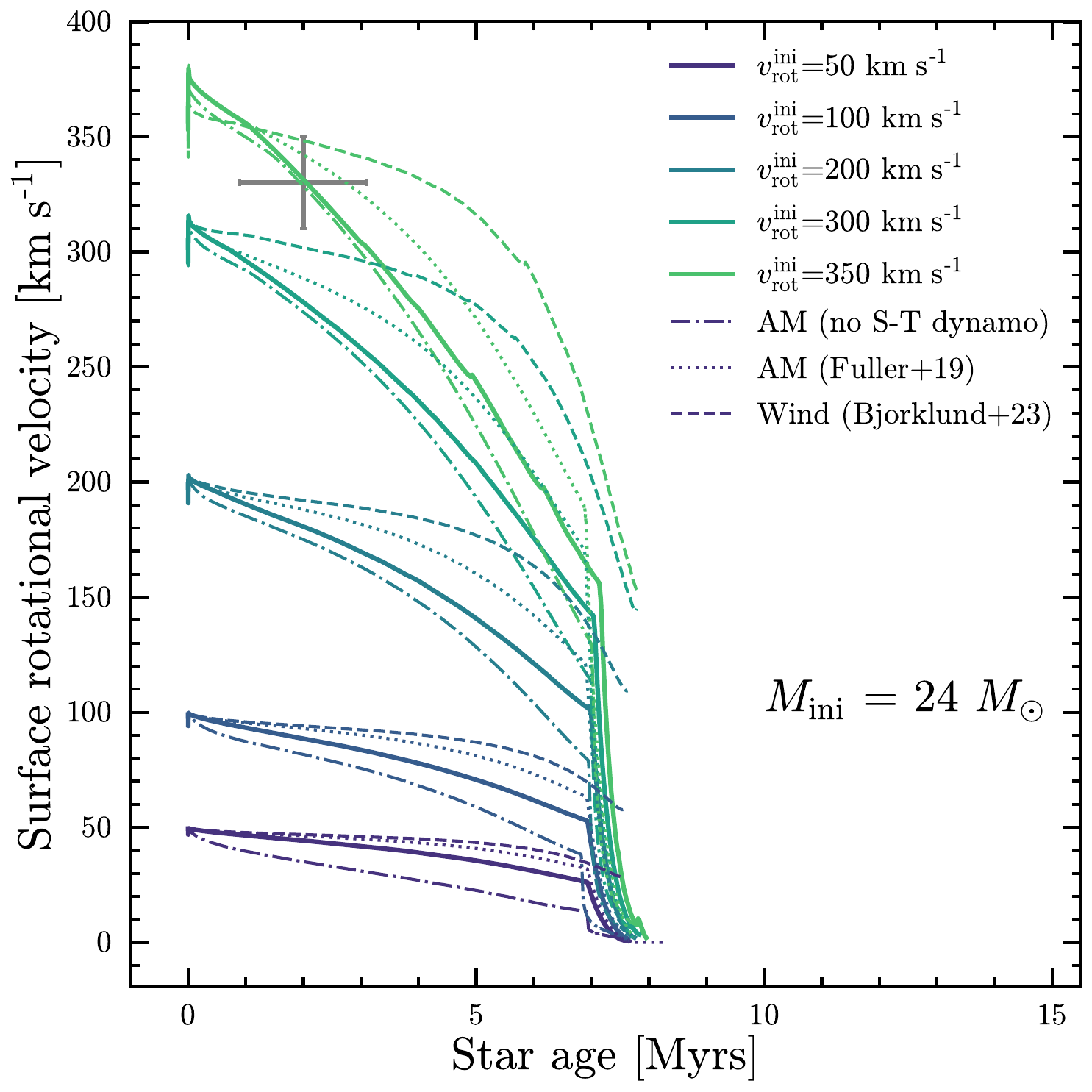}
  \caption{Evolution of equatorial surface rotational velocity of a
    single 24 $M_{\odot}$ star for different values of initial
    rotational velocity by assuming the formalism of the \citet{Vink_2001} stellar wind model (solid lines) and the \citet{Bjorklund_2023} wind model (dashed lines). In addition, tracks without the Spruit - Tayler dynamo mechanism (dotted-dashed lines) and the tracks assumed \citet{fuller:19} AM transport (dotted lines) are added for reference.
 The tracks represent only the main-sequence
    stage up to terminal-age evolutionary phase. The error bars
    represent the observed position of HD\,46485, whose primary has comparable mass.}
  \label{fig:v_surf_one_star}
\end{figure}

We start our modeling from the investigation of the rotational
properties of a 24\,$M_{\odot}$ star evolving as a single star. First, we
show for illustrative purposes in \Figref{fig:v_surf_one_star} the MS
evolution of the surface equatorial rotational velocity
($v_\mathrm{rot}$) with varying initial values
$v_\mathrm{rot}^\mathrm{ini}$ at solar metallicity ($Z=0.02$).

Our choice of initial mass roughly corresponds to the primary
  mass of HD\,46485 which is the largest one in our sample, thus the one
most affected by wind spin-down. Indeed, in
\Figref{fig:v_surf_one_star} we can see that, regardless of the
initial rotation, the star slows down to
$v_\mathrm{rot}\sim 5$\,\kms~ within $\sim$8\,Myrs \citep[solid lines,][]{Vink_2001}. The evolutionary tracks stop at the
terminal age main-sequence (TAMS, defined by the central mass fraction
hydrogen decreasing below of $X(^{1}\mathrm{H})<0.01$). We can
notice around $\sim7$\,Myr the large decrease in rotation and mass caused by
  the increase in mass loss rate due to the ``bi-stability jump''
(e.g.,\citealt{Vink_2001} but see also \citealt{bjorklund:21})
which also removes a large portion of the available angular
momentum of the star. The physical reason that causes
such significant changes in mass loss rate is the recombination of
  iron ions at certain effective temperatures ($\sim$ 21 000 and 10
000 K) which dramatically increases the opacity and thus the
  effectiveness of the line-driven wind in removing mass, and as a result it reduces the angular momentum.
This rotation braking  occurs mostly for O-type stars at late stages of the MS. However, the stellar wind model based on \citet{Bjorklund_2023} study does not predict the existence of the ``bi-stability jump'' (see dashed lines in \Figref{fig:v_surf_one_star}) because it is based on a different radiation-driven mass loss prescriptions.

Thus, there is still no general consensus regarding
which mass loss prescriptions \citep{smith:14, renzo:17,
  Keszthelyi_2017, bjorklund:21, Bjorklund_2023} and which model of
internal angular momentum transport \citep[e.g.,][]{fuller:19,
  denhartogh:20, Gormaz_2023} are more suitable for massive OB stars.
Consequently, there is a systematic theoretical uncertainty
  in the steepness of $v_\mathrm{rot}$ drop and
the size and duration of the bi-stability jump can vary depending on
which underlying physics is adopted (e.g., mixing efficiencies,
mass loss rates, and magnetic braking). Another effect
  contributing to the spin down of the star is the evolutionary
expansion of the envelope, controlled primarily by the change in
  mean molecular weight \citep[e.g.,][]{xin:22} and thus
  what fraction of the hydrogen mass is burned in the convective
  core and ultimately the convective boundary mixing \cite[e.g.,][]{Brott_2011, johnston:21}.

Because of conservation of AM, as the radius increases the surface rotational frequency has to decrease -- with angular momentum transport and wind losses adding complications to this simplified argument.
However, since all our targets are located close to ZAMS, there is little time for this to happen.
Without the Spruit - Tayler dynamo mechanism, AM transport less efficient, thus the stars spin down faster because of the wind (see dashed-doted lines in \Figref{fig:v_surf_one_star}).
Conversely, assuming the more efficient AM transport from \citet{fuller:19}, the star is spin down more slowly (see dotted lines in \Figref{fig:v_surf_one_star}).
This illustrates that while a star is evolving, the core contracts and spins up and the resulting surface spin down rate directly depends on the assumed efficiency of AM transport.
However, the difference in surface rotational velocity caused by different AM transport mechanisms is small within the early MS stage.

Another factor that influences the rotation rate is the assumed wind mass loss rate.
A good illustration of how wind underlying theory is affecting the evolution of $v_\mathrm{rot}$ can be seen comparing tracks computed with \citet{Bjorklund_2023} (dashed), where the increase in wind mass loss because of bi-stability jump(s) is absent, and those computed with \citet{Vink_2001} (solid lines in \Figref{fig:v_surf_one_star}).
The overall mass loss rate in the former is lower, allowing the stars to retain more AM and a faster surface rotation rate throughout the main sequence.
The difference between \citeauthor{Bjorklund_2023} and \citeauthor{Vink_2001} mass loss rates lies in the new prescription of radiative acceleration which covers a large range of atmosphere layers including supersonic wind outflow which is considered in \citet{Bjorklund_2023} approach.
The mass loss rate is then derived as a scaling factor of fundamental stellar parameters based on a large grid of models.
The resulting mass loss rate appears to be weaker than the previous estimates and as a result the $v_\mathrm{rot}^\mathrm{ini}$ may remain high for a significant time (up to the end of the MS).

However, differences in $v_\mathrm{rot}$
behavior become significant only near the end of the MS, a stage far from
current evolution properties of our targets (see \Figref{fig:hr_plot_paper2023}). The results we presented in
\Figref{fig:v_surf_one_star} can therefore be adopted as a typical
$v_\mathrm{rot}$ behavior for the young OB star domain. The resolution
test for a given star model is presented in
\Figref{fig:resolution_test_one_star}.

\subsection{Tidal interaction}

\begin{figure}[!ht]
  \centering
  \script{plot_save_mesa_comparison_panels_fin.py}
  \includegraphics[width=0.50\textwidth]{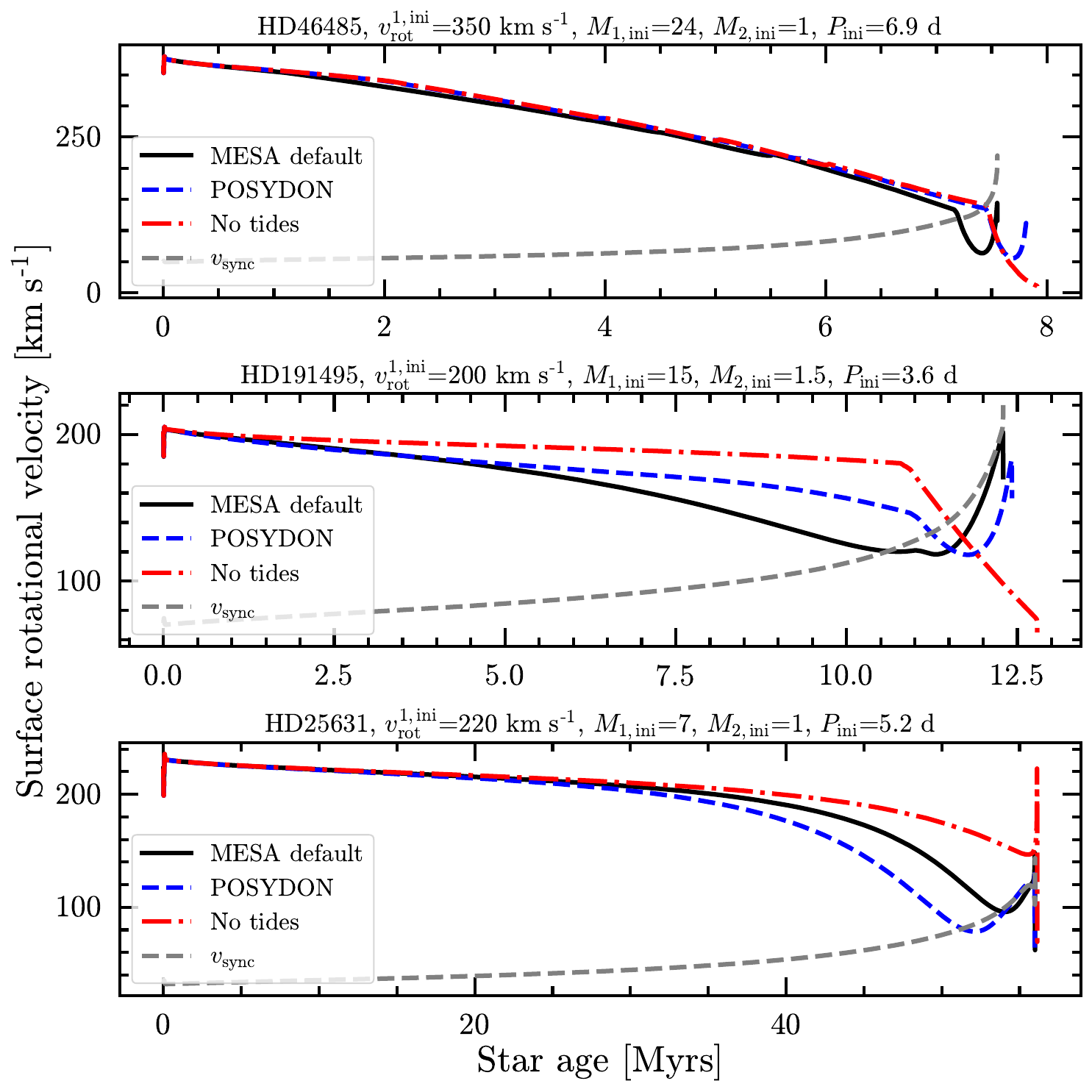}
  \caption{Evolution of equatorial and synchronization surface rotational velocity for our
    three program systems by taking into account different tidal
    synchronization prescriptions. "MESA default" refers to the
    radiative envelope tidal synchronization (\texttt{"Hut$\_$rad"}
    prescription), and "POSYDON" to the structure dependent
    prescription. We emphasize that different scales for the axes of
      each panel are used due to the different rotation rates, stellar
      masses, and thus lifetimes of the systems.}
  \label{fig:v_surf}
\end{figure}

The next step is to consider stars evolving in a binary system and add the effect of tides on the rotational velocity
and how they affect the evolution of the surface rotational
  velocity of our systems, shown in \Figref{fig:v_surf}. We
start with binaries with architectures comparable to the ones observed
today, with fast-rotating super-synchronous primaries and initial rotation of secondaries set to $v_\mathrm{rot}^\mathrm{2,ini}$ = 1 \kms.
We run three binary systems with ($M_\mathrm{1,ini}$, $M_\mathrm{2,ini}$,
$P_\mathrm{ini}$, $v_\mathrm{rot}^\mathrm{1,ini}$ ) = (24 $M_{\odot}$, 1 $M_{\odot}$, 6.9 d, 350 \kms), (15 $M_{\odot}$, 1.5 $M_{\odot}$,  3.6 d, 200 \kms), (7 $M_{\odot}$, 1 $M_{\odot}$, 5.2 d, 220 \kms) for HD\,46485, HD\,191495, HD\,25631, respectively (based on the values presented in \Tabref{table:sb1_fast}).
By default we used the \citet{Vink_2001} mass loss prescription and we evolved the systems beyond the initial onset of mass transfer.

To explore the large uncertainties in modeling of tidal interactions
\citep[e.g.,][]{zahn:75, Zahn_1977, Qin_2018, preece:22, fuller:22}, we use three different
available tidal synchronization algorithms. Namely, we used
no tides at all (red dot-dashed lines in \Figref{fig:v_surf}), the tidal synchronization prescription for radiative envelopes
(\texttt{sync\_type = "Hut\_rad"}, aka "MESA default", see black solid lines) based on \citet{Hut_1980} and
\citet{Hurley_2002} as well as the structure-dependent
prescription (\texttt{sync\_type = "structure\_dependent"})
implemented in the POSYDON code \citep[blue dashed lines, based on][]{Qin_2018, Fragos_2023}.

\begin{figure}[!ht]
  \centering
  \script{plot_save_mesa_comparison_panels_acc_fin.py}
  \includegraphics[width=0.50\textwidth]{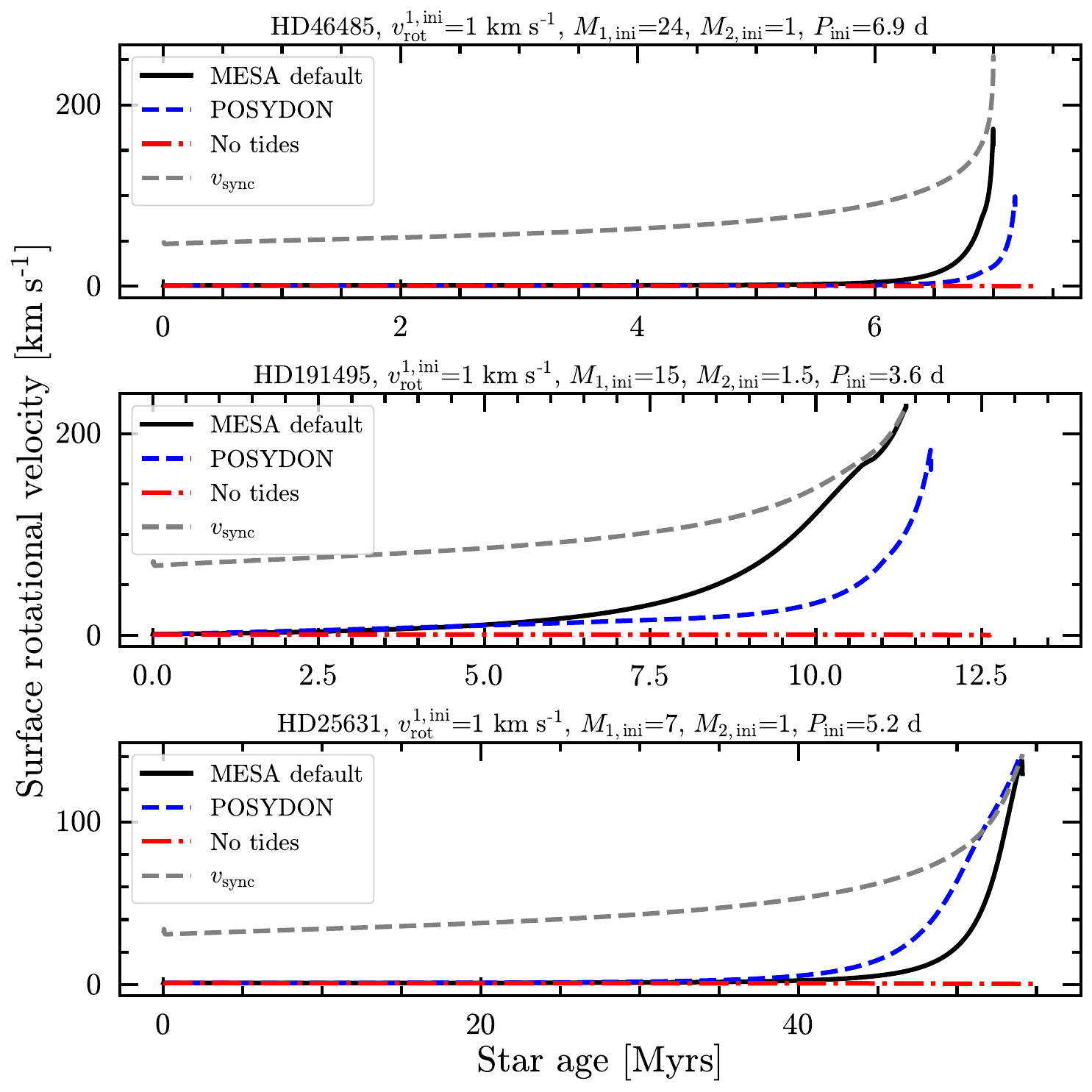}
  \caption{Same as \Figref{fig:v_surf} but assuming $v_\mathrm{rot}^\mathrm{1,ini}$ = 1 \kms. The gray dashed lines represent the evolution of synchronization rotational velocity for each system.}
  \label{fig:v_surf_acceleration}
\end{figure}

The main difference between these prescriptions is the different treatment of the synchronization timescale for radiative and convective layers depending on the
current evolutionary stage of the star.
 In POSYDON, the structure-dependent prescription checks the shortest synchronization timescale between the dynamical tides (same as \texttt{sync\_type = "Hut\_rad"}) and equilibrium tides based on the existing convective layers (similar \texttt{sync\_type = "Hut\_conv"} but not limited to a deep convective envelope) and applies it for each layer of the star, while in "MESA default" for each layer of the star the code takes into account only one dynamical timescale (\texttt{sync\_type = "Hut\_rad"}).
It is important because the mechanism of the dissipation processes of the tidal kinetic energy is different for the convective and radiative envelopes (equilibrium and dynamical tides respectively).
 Also, the code takes into account the different distribution of mass in the star depending on these
layers.

We modeled the evolution of equatorial rotation velocity for
our sample of binaries and the results are presented in \Figref{fig:v_surf}\footnote{Resolution test for the model of the HD\,46485 system is presented in \Figref{fig:resolution_test_tides}.}.
We should note that \citet{Hurley_2002} tidal synchronization prescriptions, which are implemented by default in the MESA (that explains the origin of our nomenclature "MESA default" in the text and figures), consist of \citet{Zahn_1977} dynamical tides solution which is adapted to \citet{Hut_1981} formalism together with the \citeauthor{Hut_1981}'s prescription for the equilibrium tides (which are dominant in the stellar convective zones).
Recently, \citet{Sciarini_2024} argued about the inconsistency in this approach which leads to over- or underestimation of tidal strengths depending on how close to the synchronizations the given system is, and as a result, it could affect the timescale of tidal synchronization.
Thus, it is necessary to be cautious about the original MESA treatment of tides within the \citeauthor{Hurley_2002} formalism.

\begin{figure}[!htbp]
  \centering
  \script{all_hr_fin.py}
  \includegraphics[width=0.50\textwidth]{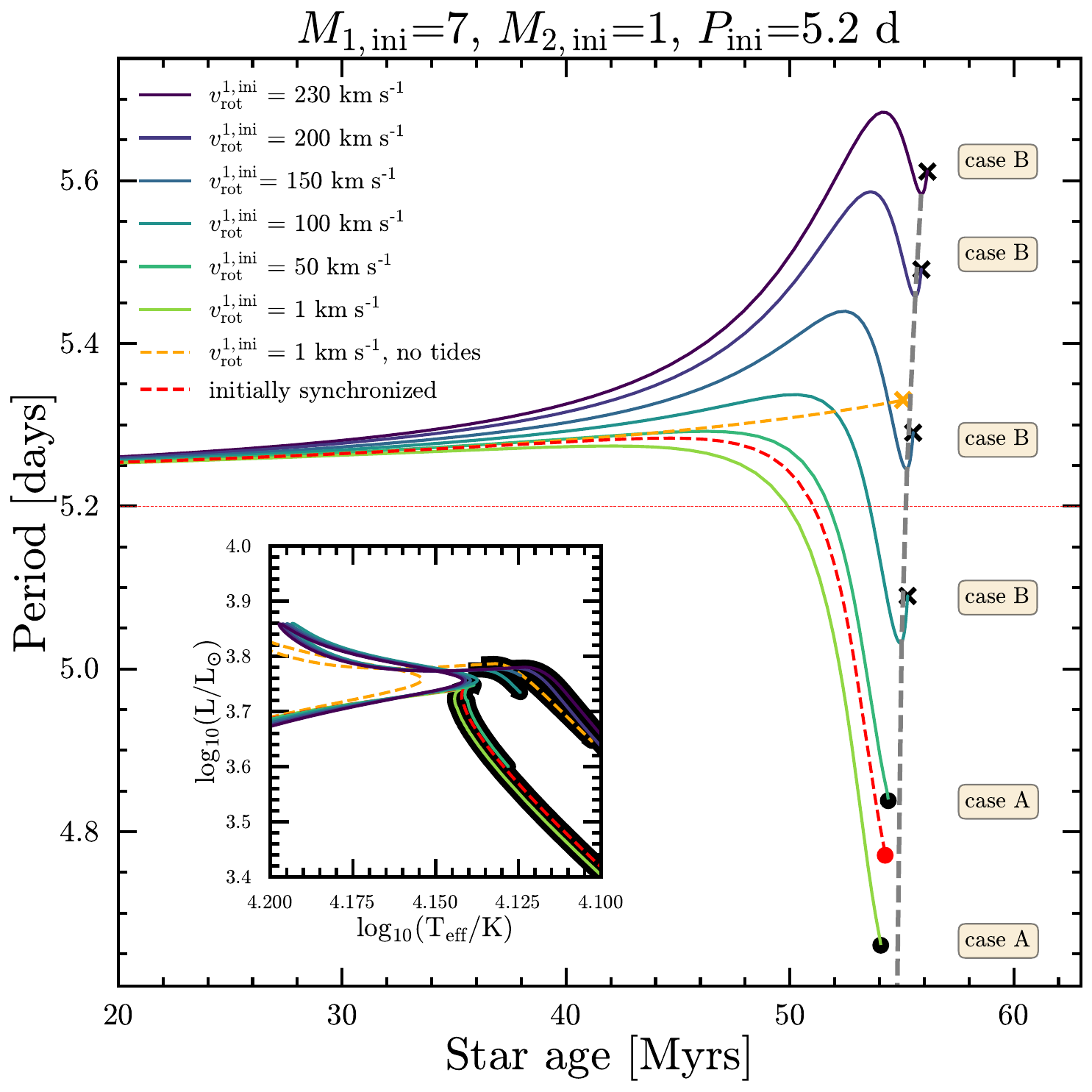}
  \caption{Evolution of the orbital period of HD\,25631 for different
    values of the initial rotational velocity (including initially synchronized rotation for a given $P_\mathrm{ini}$). The tracks stop at the
    onset of Roche lobe overflow of the primary star. When
      this occurs during its main-sequence (case A) we mark it by
      dots, conversely, if this happens after TAMS (gray dashed line)
      in a case B, we mark it with crosses. For reference, the inset shows the HR diagram around
    TAMS, where the beginning of the RLOF phases are marked by
      thick black outlines. The red horizontal line visualizes the initial value of the orbital period. In these simulations we used the \texttt{sync\_type = "Hut\_rad"} tidal prescription.}
  \label{fig:v_surf_hr}
\end{figure}

We adopted the present \vsini~ as the initial value of equatorial
rotation. In this way, we can estimate how long such a high rotational
regime can exist and how the tides are affecting it (although see
below for a numerical experiment varying the initial rotational
velocity, cf.~\Figref{fig:v_surf_hr}). As we can see
from \Figref{fig:v_surf}, the tides significantly affect the systems
with a less extreme mass ratio (i.e., HD\,25631 and HD\,191495
shown in the bottom and middle panels of \Figref{fig:v_surf}).
As expected, the tides are mainly spinning down these stars and widening the orbit.
Since all of the investigated systems have super-synchronous rotation and short-periods, the tidal interaction can not be a reason of the
observed high \vsini.

However, at the very end of the MS
(see \Figref{fig:v_surf}), we can see a quick spin up for each of
the stars, caused by tidal interactions. At that moment, the stellar
radius is increasing approximately by a factor of two, and with
such short orbital periods, the tidal interaction starts to
synchronize quickly orbital and rotational velocities. In the case of
HD\,191495 (middle panel of \Figref{fig:v_surf}), the rotational
velocity without tides drops down at the "bi-stability jump" stage,
while with tides the star is spinning up to synchronization velocity
($v_\mathrm{sync}$) when the radius of the star reaches 14
$R_{\odot}$ ($v_\mathrm{sync}$ for this radius and the
given period is $\sim$ 200 \kms).

It is important to note that in some cases (HD\,25631 and HD\,191495) close to the TAMS, we have a non-negligible difference in resulting $v_\mathrm{rot}$ between "MESA default" (\texttt{sync\_type = "Hut\_rad"}) and POSYDON tides prescriptions.
That is caused by the different treatment of the dynamical tides in the radiative envelope which are assumed in these prescriptions.
In the POSYDON tides formalism (\texttt{sync\_type = "structure\_dependent"}), the synchronization timescale of dynamical tides is more sensitive to the structure of the star via tidal coupling parameter \citep[often called $E_{2}$, see Section 3 in][]{Qin_2018} which depends on the radius of the convective core, while in \citet{Hut_1980,Hurley_2002} formalism these types of tides just depend on the entire mass of the star (see \Appref{sec:e2_def} for details).
However, the main difference of the POSYDON prescription is that it takes the shortest of the calculated dynamical and equilibrium synchronization timescales, with the latter expected to become important for giant stars or even massive stars with significant convective layers during their MS.
We detected the maximum difference by a factor of three between the resulting synchronization times in the case of our systems.
Thus, for the calculation of the tidal synchronization timescale of massive stars, we cannot neglect the sizes of convective and radiative envelopes.
Obviously, in order to be more precise regarding the effect of tides on the stellar spin, a more realistic model of tidal dissipation is needed \citep[especially while investigating the spin distribution of black holes, see e.g.][]{Bavera_2020,Ma_2023}.

To investigate how $v_\mathrm{sync}$ is varying while the systems are evolving we repeated the same simulations with varying tidal
 prescriptions as before but assuming the initial $v_\mathrm{rot}$ equal to 1 \kms.
In this way, we can estimate the spin up timescale for each of our systems.
The results of these simulations are presented in \Figref{fig:v_surf_acceleration}, as we can see the systems reach the $v_\mathrm{sync}$ only at the end of MS.
The synchronization velocity is calculated as $v_\mathrm{sync} = 2  \pi  R / P $. Where, $R$ and $P$ are the radius of the primary component and orbital period of the system assuming \texttt{sync\_type = "Hut\_rad"} tidal synchronization prescription.
In order to see the effect of tidal acceleration which is free from the stellar radii expansion, we also checked the behavior of surface average angular velocity ($\omega$, see \Figref{fig:v_omega}), which also reached the synchronization values at TAMS.

This analysis shows that tidal interactions alone are not able to spin up our primaries to the present values of $v_\mathrm{rot}$.
In addition, we also checked how fast the circularization of the systems could happen (see \Figref{fig:v_ecc}), it appears that for such short-period systems, the circularization timescale could be as long the entire MS duration.
Thus, with the apparent young age of our stars, tides cannot have changed a lot the eccentricity and rotation of the systems.
These simulations also allow us to evaluate another evolutionary scenario: the systems are originally wide binaries observed post-common envelope ejection and the spin is a product of post-common envelope spin-up.
However, this scenario is a priory problematical because (\emph{i}) we know the unseen companion to be cooler than the observed stars and (\emph{ii}) tides cannot spin up to super-synchronous rotation, as we illustrate here the timescale for the tidal spin up.

Interestingly, the starting moment of mass transfer is delayed when one considers both high rotation velocity and tidal interactions.
To illustrate this, we model the HD\,25631
system that has the maximum tidal effect, i.e., the one with the largest
mass-ratio, with different initial rotational velocities. In
\Figref{fig:v_surf_hr} we show how the orbital period varies depending
on the initial rotational velocity up to the moment the donor reaches
the RLOF phase. As we can see, with $v_\mathrm{rot}^\mathrm{ini}$ $>$ 100
\kms~ the mass transfer starts after the main-sequence phase (i.e.
case B in \citealt{kippenhahn:67} notation). While it happens before (case A) for slower rotators. We
also plot the HR diagram for reference and indicate the TAMS by a gray
dashed line. The delaying of the binary mass transfer occurs because, with a
larger rotational velocity, the tides make the orbit wider, therefore it takes more time to reach the RLOF stage.
This effect of switching mass transfer from case A to early case B is not so prominent in the case of low mass ratios as we show in the example of our system.
However, it may play an important role in subsequent binary evolution
in case of larger mass ratios (the mass transfer rate will be larger and unstable as the donor enters rapid phases of evolution).
In the post-main sequence evolution (not shown in \Figref{fig:v_surf}), the primary fills its
Roche lobe, starting a phase of mass transfer which results in spin down of the donor star which loses mass and angular momentum.
Moreover, a significant rotation of the donor onset the mass-transfer affects the gas trajectory, and as a result it could affect the mass transfer efficiency and the presence of the disk around the accretor \citep[see,][]{Hendriks_2023}.
Also, we calculated the evolutionary track without any tidal interactions (see orange dashed line in \Figref{fig:v_surf_hr}) that represents the widening of the orbit just from the effect of mass loss.
In this way, we can see that depending on the initial rotational velocity, the effect of tidal interaction is bigger than the effect of the mass loss.
All these models (as all in the present work) do not include the initial rotational synchronization to the initial period of the system (as all of our systems have super-synchronous rotation).
However, we assumed such a scenario (red dashed line in \Figref{fig:v_surf_hr}, that leads to the case A mass transfer), as the reference to the rest of the mentioned tracks.

\begin{figure*}[!hbpt]
  \centering
  \script{plot_save_mesa_ce.py}
  \includegraphics[width=0.33\textwidth]{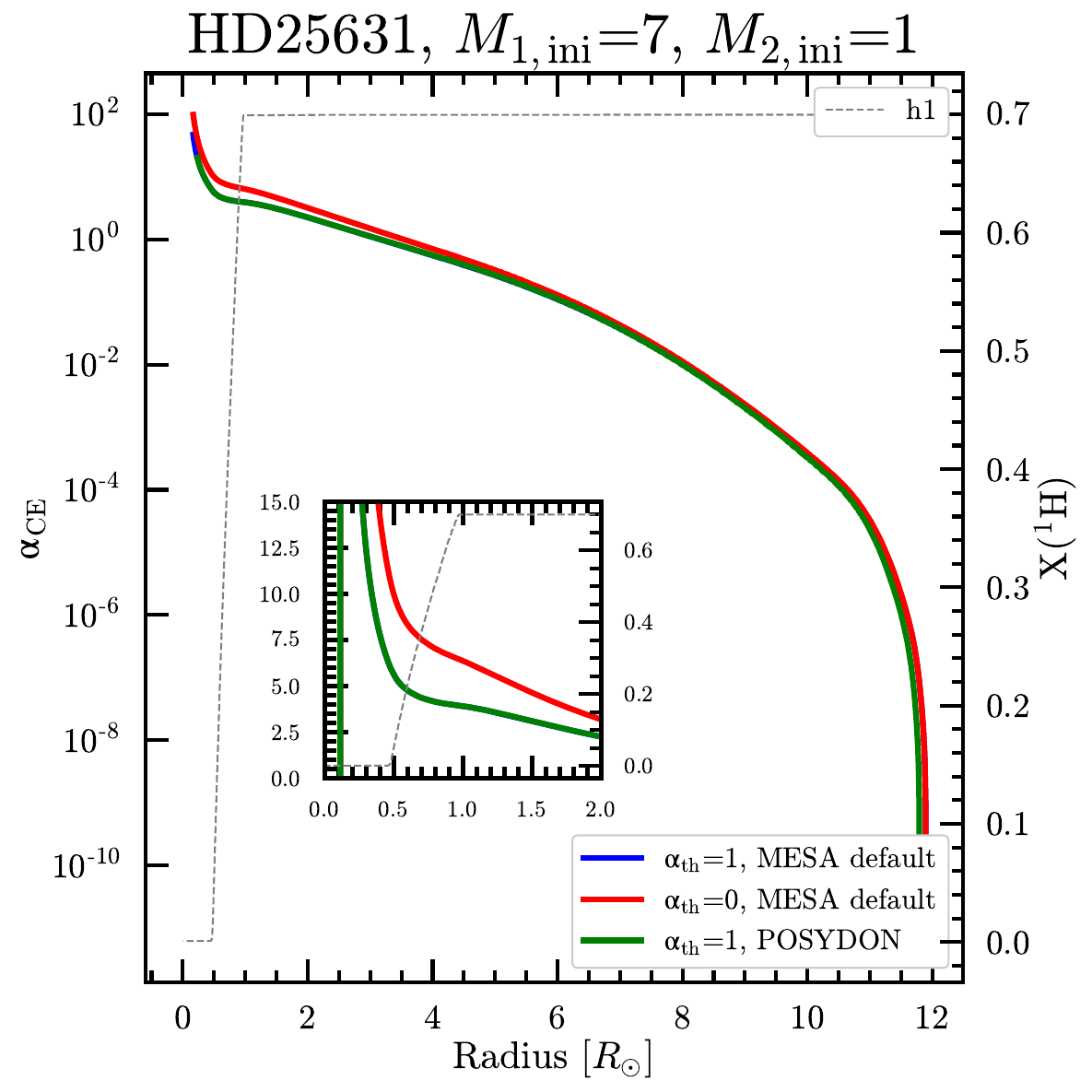}
  \includegraphics[width=0.33\textwidth]{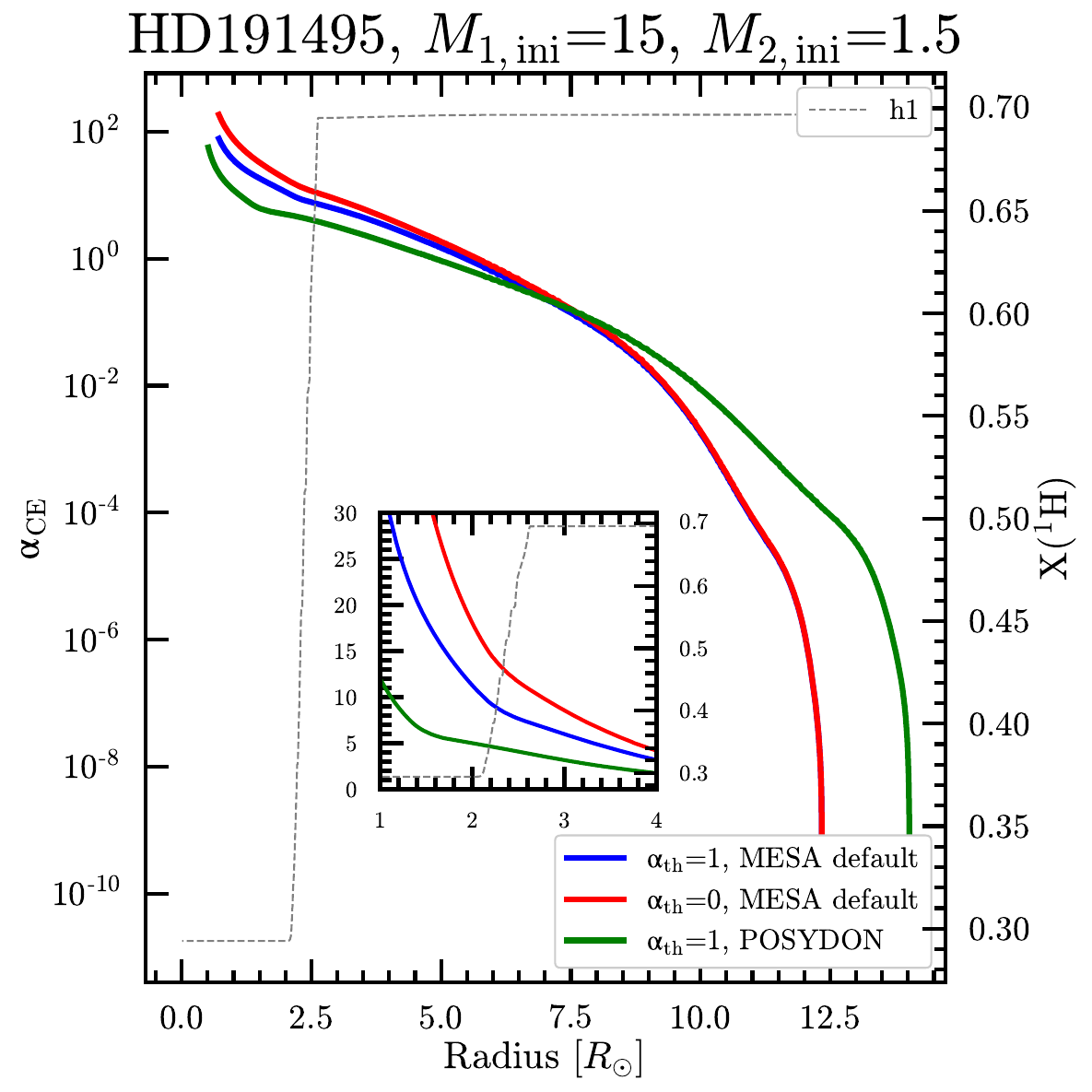}
  \includegraphics[width=0.33\textwidth]{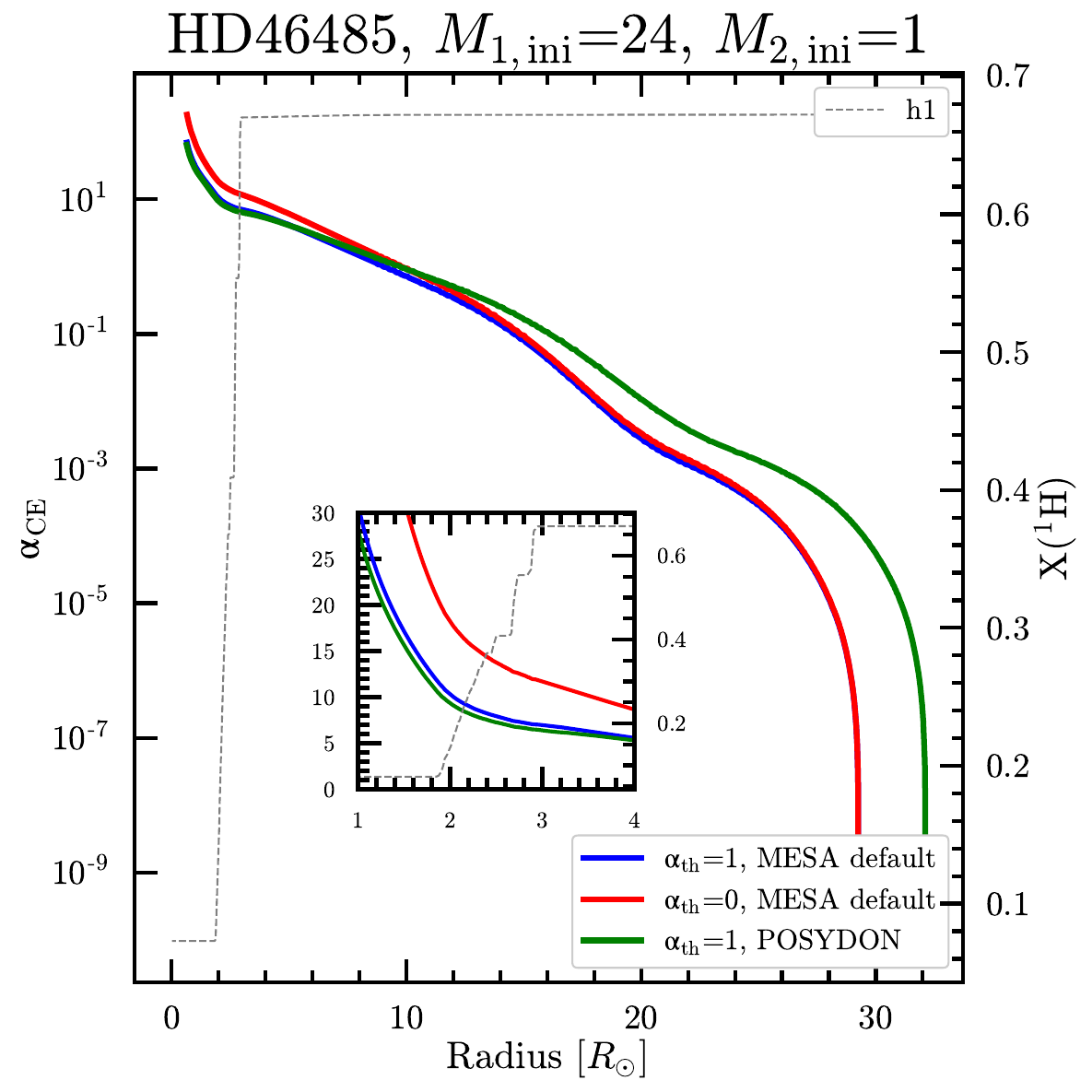}
  \caption{Efficiency of the CE process $\alpha_{CE}$, with a
    different fraction of thermal energy $\alpha_{th}$ for our three
    targets (see text for details). In order to define the zone of the
    stellar core we indicated the mass fraction of $X(^{1}\mathrm{H})$
    by the gray line, linked to the right y-axis.}
  \label{fig:hr_alpha}
\end{figure*}

Our simulations lead us to the conclusion that the spin-down timescale is comparable to the entire main-sequence lifetime of each star.
For example, for 7 $M_{\odot}$ and 15 $M_{\odot}$ primaries (HD\,25631 and HD\,191495, respectively), the $v_\mathrm{rot}$ only drops by 4 $\%$ and 6 $\%$ from the initial value at the half of the main-sequence duration.
In the case of the more massive (24 $M_{\odot}$) star HD\,46485, the
drop in $v_\mathrm{rot}$ is happening quicker (by 22 $\%$ at half of main-sequence duration, assuming \texttt{"Dutch"} mass loss regime) because of the
stronger mass loss rate for stars of larger initial mass.
In addition to this, we show that tidal synchronization can spin up the star only at the end of the main-sequence when the size of the star is significantly larger than at ZAMS.
In view of the young age of all studied systems \citep[$<$ 10 Myrs,
see \Tabref{table:sb1_fast} and][]{Naze_2023_rot} and our
demonstration that the initial rotation can remain for a significant
duration of the main-sequence (cf.~\Figref{fig:v_surf_one_star}), we suggest that the observed fast rotation
of these systems can be primordial.
With the current short orbital periods and extreme mass ratios, tidal
synchronization only plays an important role during the latest stages
of the main-sequence.
This is in agreement with other studies of synchronization timescale of Algol systems that appeared to be longer and weaker than accretor's spinning-up timescale due to mass transfer \citep[see, e.g.][]{Deschamps_2013}.

\subsection{Future of the systems}

We evolved the models presented in \Secref{sec:res_primordial}
  until the onset of binary mass-transfer, occurring right at the end
  of the main-sequence or just after, depending on the initial
  rotation. Taking into account the rather extreme mass-ratios of these systems, it is natural to consider
the common envelope (CE) scenario when investigating the future of our
systems \citep{paczynski:1976,claeys:14,renzo:21gwce}.

The largest mass ratio $q\lesssim0.15$ (see \Tabref{table:sb1_fast})
  is indeed much smaller than the  typical minimum threshold for stability \citep[e.g.,
  $q\geq q_\mathrm{crit}\simeq0.25-0.625$ for main-sequence
  interactions,][]{claeys:14}. Thus, we assume that the mass-transfer
  event will become dynamically unstable and result in a common
  envelope (CE) event \citep{paczynski:1976, ivanova:2013, ivanova:2020,
    renzo:21gwce}.

We calculate the $\alpha_\mathrm{CE}$ parameter
  \citep{webbink:1984} to discuss the chance of a successful ejection.
  This parameter measures the efficiency of the CE process in using
  orbital energy to lift the shared envelope,

\begin{equation}
  \label{eq:alpha_CE}
  \alpha_\mathrm{CE} \equiv \alpha_\mathrm{CE}(r) = \frac{E_\mathrm{bind}(r)}{\left(-\frac{GM_1M_2}{2a_\mathrm{pre-CE}}+ \frac{Gm_1(r) M_2}{2r}\right)} \ \ ,
\end{equation}
where $E_\mathrm{bind}(r)$ is the binding energy profile (assumed
  to be ``frozen'' during the CE, thus we neglect the thermal and dynamical response of the envelope to the inspiral), $G$ the gravitational constant, $m_1\equiv m_1(r)$ is the donor mass remaining inside
$r$ such that $m_1(r=R_1)=M_1$, $M_2$ is the companion mass,
$a_\mathrm{pre-CE}$ is the binary separation at the moment of TAMS \footnote{Among our systems only the HD\,25631 reaches the TAMS before the mass transfer would start (case B), thus for the calculation of $\alpha_\mathrm{CE}$ we used the required parameters at the moment of TAMS for HD\,25631, and at the moment right before the mass transfer (case A)  is started for the other systems.} (i.e., $\sim$ 45.0 $R_{\odot}$, $\sim$ 25.7 $R_{\odot}$, $\sim$ 26.3 $R_{\odot}$ for HD\,46485, HD\,191495, and HD\,25631 respectively).
Formally, conservation of energy requires
  $\alpha_\mathrm{CE}<1$ \citep{Iaconi_2019} although larger
  values are routinely considered to mimic the inclusion of energy
  sources other than the orbit in the common-envelope ejection
  process \citep{Han_1995, ivanova:02, demarco:11, zorotovic:14}.

To determine $E_\mathrm{bind}(r)$, we integrate the internal structure of our stellar models from the surface inward
  \citep[e.g.,][]{Han_1995, Dewi_2000, renzo:23}:
\begin{equation}
  \label{eq:BE}
E_\mathrm{bind}(m_1(r)) = - \int_{m_1}^{M_1}\,dm'\left( -\frac{G m'}{r(m')}+\alpha_\mathrm{th}u(m')\right) \ \ ,
\end{equation}
where $\alpha_\mathrm{th}$ is a tunable parameter to include
  ($\alpha_\mathrm{th}=1$) or exclude ($\alpha_\mathrm{th}=0$) the
  contribution of the internal energy $u$. Higher $\alpha_\mathrm{th}$
  result in a lower value of $\alpha_\mathrm{CE}$ which then result in an ejection of
  the envelope down to a given depth $r$,
  corresponding to an optimistic ejection case (see, e.g.,
  \citealt{klencki:21}). Since our models predict the interaction to
  be very close to the end of the MS, we focus on the results derived
at the latest steps of our simulations near the TAMS. This
  neglects the impact of the initial RLOF
  phase preceding the CE on the stellar structure and binding energy \citep[e.g.,][]{ivanova:2020, renzo:21gwce,
    blagorodnova:21}. However, this impacts mostly the outer layers of
the star and we expect it to be insufficient to change our
conclusions.

\Figref{fig:hr_alpha} presents the calculated $\alpha_\mathrm{CE}$ for
primaries of our three systems as a function of radius inside the star
and for different tidal prescriptions including and excluding the
  internal energy contribution. It
also shows the 
hydrogen mass fraction profile
($X(^{1}\mathrm{H})$) which decreases at the core-envelope boundary, see inset panel.

If, as we expect, the future mass transfer in the
  primordial scenario does result in a common envelope,
$\alpha_\mathrm{CE}$ needs to be lower than one for the
  companion to remove the entire envelope down to the
  hydrogen-depleted layers using only the orbital energy. Our simulations (\Figref{fig:hr_alpha}) indicate results of $\alpha_\mathrm{CE}$ estimates to be 3
-- 6 for HD\,25631, 5 -- 13 for HD\,191495, and 9 -- 15 HD\,46485.
This strongly suggests that the envelope will not be successfully ejected, hence the
systems will most probably merge in the future.

\section{Spin-up -- Post-interaction scenario}
\label{sec:post-interaction}

Binaries can also produce fast-rotating stars during mass-transfer episodes. As
the donor overfills its Roche lobe, the transferred mass carries
angular momentum and, if successfully accreted by the companion star,
deposits the angular momentum at its surface.
The accretor gains a significant angular momentum and spin is
increasing very quickly up to the critical rotation \citep{packet:81, Langer_2003, Renzo_2021}.

 In this scenario, the observed star in the SB1 binaries under consideration has previously been spun up by accretion and the
  unseen companion is the stripped donor \citep[e.g.,][]{Gotberg_2023, Drout_2023} or its remnant.
Given the difficulty of identifying stripped star companions, we still consider the binary mass transfer scenario to illustrate its impact on the orbital architecture, which would have made us disfavor this scenario even in absence of a detectable reflection effect.
Thus, despite the fact that all three investigated systems are pre-interacting we still aim to model them within the post-interacting paradigm which could be useful for future studies.
However, for this scenario to be valid for our targets, it needs to reproduce
all parameters of the systems (mass-ratio, periods, age, etc), not only fast rotation of the mass gainer.

\subsection{Method}

We use the binary module in MESA to evolve both stars in each binary system. To test whether post-interaction binary scenario is applicable to our systems, we
built a grid of binaries with various masses of donor star ($M_\mathrm{don}$ = [30, 20, 10] $M_{\odot}$) and accretor's masses ($M_\mathrm{acc}$ = [20, 15, 10, 7] $M_{\odot}$) that give mass ratios of $Q$ =
$M_\mathrm{acc}$/$M_\mathrm{don}$ = [0.75, 0.7, 0.65, 0.5, 0.33]. 
For these systems we use initial periods of $P_\mathrm{ini}$ = [1.25, 3, 5, 10, 50, 70] days. Given the small observed
  eccentricity (\Tabref{table:sb1_fast}), and the significant circularization timescale (\Appref{sec:angular_vel}), we assume our binaries
  to be initially circular ($e=0$). Our setup is closely based
  on \cite{Renzo_2021, renzo:23}, and we summarize here the main
  physical and numerical choices regarding initial rotation and mass transfer.
  We refer to \Appref{sec:MESA_setup} and the previous section for
  further information.

We model Roche lobe overflow (RLOF) allowing
  for optically thick outflows \citep{Kolb_1990}. Mass
transfer is conservative up to the point when the
accretor reaches the critical rotation
$\omega_\mathrm{crit}=\sqrt{(1-L/L_\mathrm{Edd})GM/R^3}$ where centrifugal force plus radiation pressure equal
  gravity at the equator, with $L_\mathrm{Edd}$ the Eddington
  luminosity, and $L$ the stellar luminosity \citep[e.g.,][]{Petrovic_2005,Heuvel_2017}. The transferred
  material carries the same specific angular momentum as the accretor
  surface, producing a relatively slow spin-up \citep{Renzo_2021}.
When the accretor's surface approaches critical
  rotation, $\omega = 0.95\omega_\mathrm{crit}$, we enhance the
  accretor's mass loss rate following \cite{langer:98}, which briefly
  stops the accretion. The mass loss spins down the stars, resulting in
  accretion resuming in a ``self-regulated'' way, moderated by the
  inward angular momentum transport \citep[e.g.,][]{Renzo_2021}. Overall, this leads
  to a mass transfer efficiency \citep[$\eta =\Delta M_\mathrm{acc}/\Delta M_\mathrm{don}$ as it has been introduced in][]{Soberman_1997} that actually varies depending on the orbital configuration i.e., in close binaries the mass transfer is more efficient (see the next subsection).

To test a lower limit on how fast the accretors may be spun up we
  initially assume rigid rotation with $v_\mathrm{rot}^\mathrm{ini} = 1$\,\kms~ at the equator for both stars, regardless of the orbital period.
Our models include tides, following the two prescriptions
  discussed in \Secref{sec:res_primordial} -- thus on a tidal
  timescale we expect the orbits to shrink slightly to spin up both
  stars.

\subsection{Results and interpretation}

As an illustrative example, we first focus on $M_\mathrm{don,ini}$ = 30 $M_{\odot}$ and
$M_\mathrm{acc,ini}$ = 20 $M_{\odot}$, corresponding to $Q=0.65$, close
to the average value of a flat mass ratio distribution between 0.1 and
1, as observed in massive OB-type binaries \citep[see][]{Vanbeveren_1981,Sana_2013}. Moreover, to test the
  viability of the accretor scenario, we aim to reproduce with an
  accretor star the highest mass case among our fast rotators (HD\,46485, $M_\mathrm{1, obs}$ = 24 $M_{\odot}$).
As mentioned before, we did explore a larger range of parameters but it is important to note that all simulations lead us to similar conclusions, hence our choice to present in detail only one case.

\begin{figure*}[!ht]
  \centering
  \script{plot_save_mesa_individual_fin1p25.py}
  \includegraphics[width=0.30\textwidth]{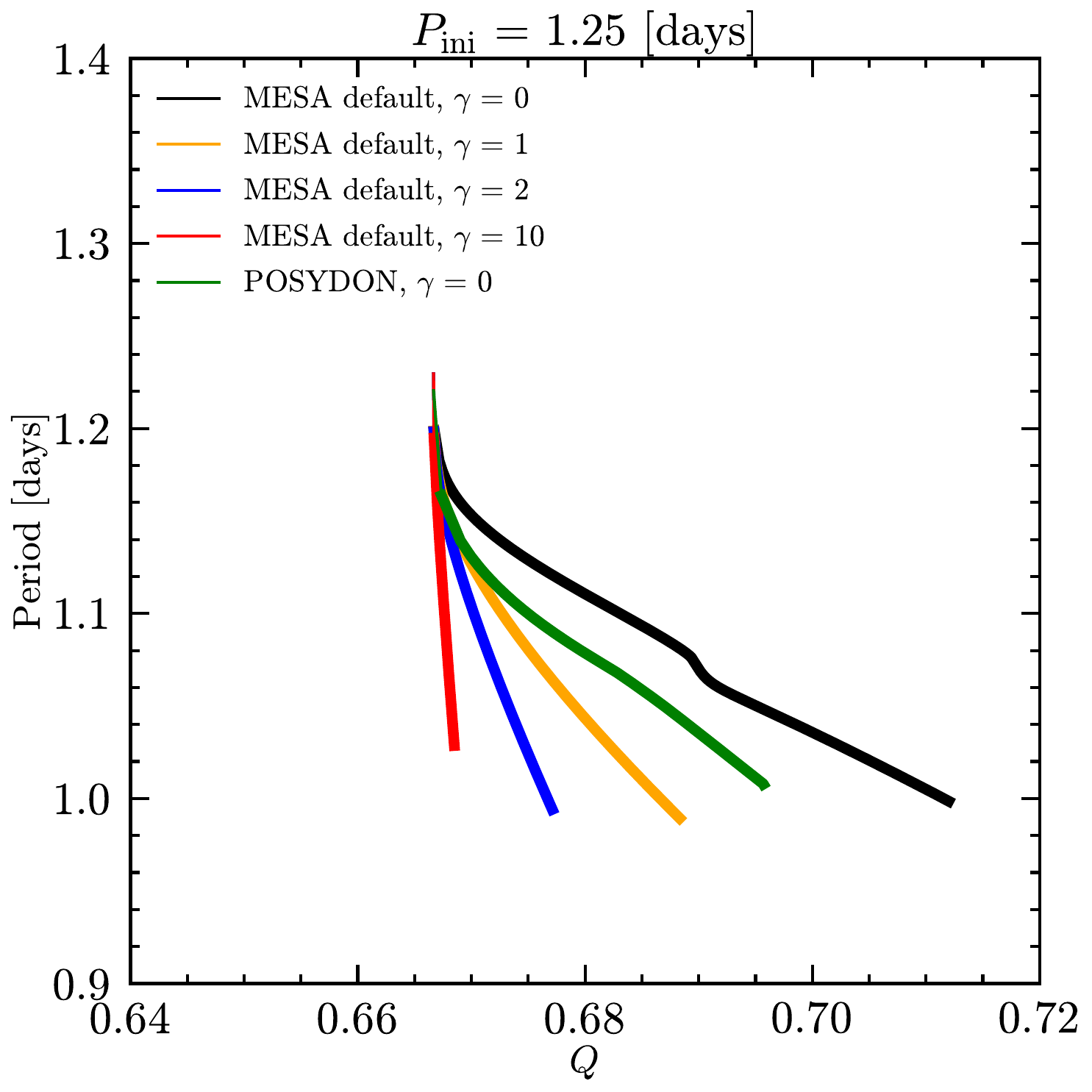}
  \includegraphics[width=0.30\textwidth]{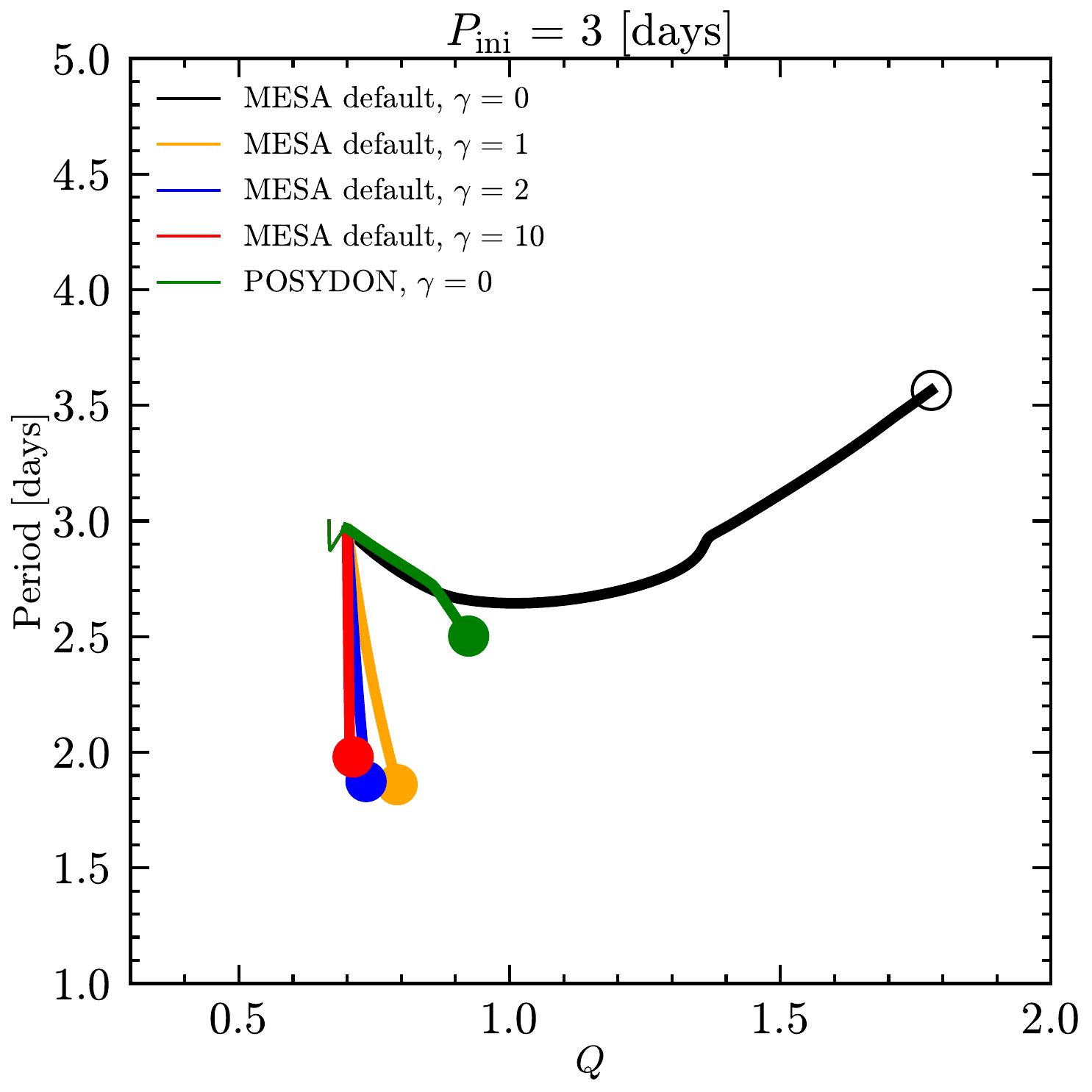}
  \includegraphics[width=0.30\textwidth]{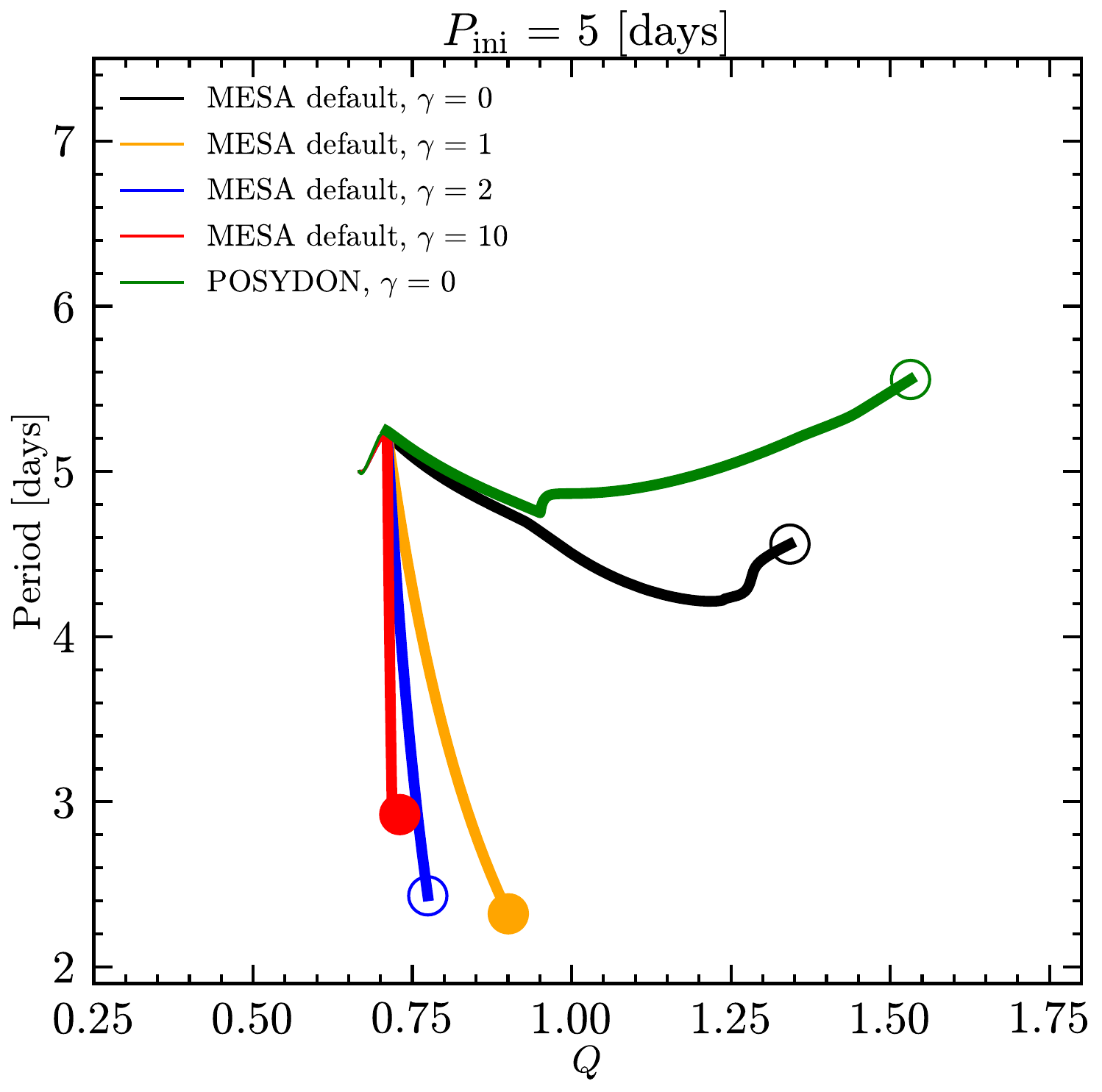}
  \includegraphics[width=0.30\textwidth]{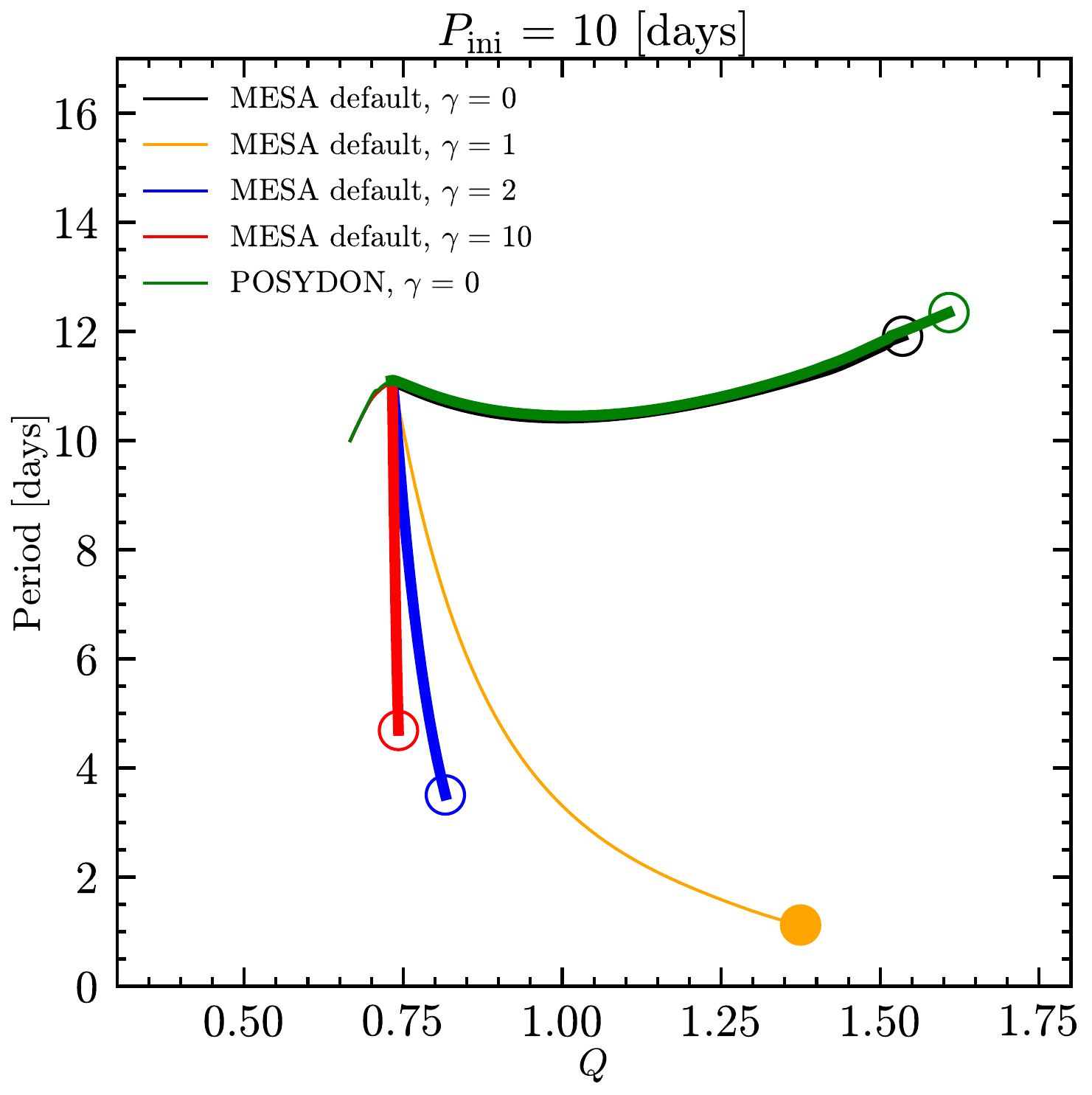}
  \includegraphics[width=0.30\textwidth]{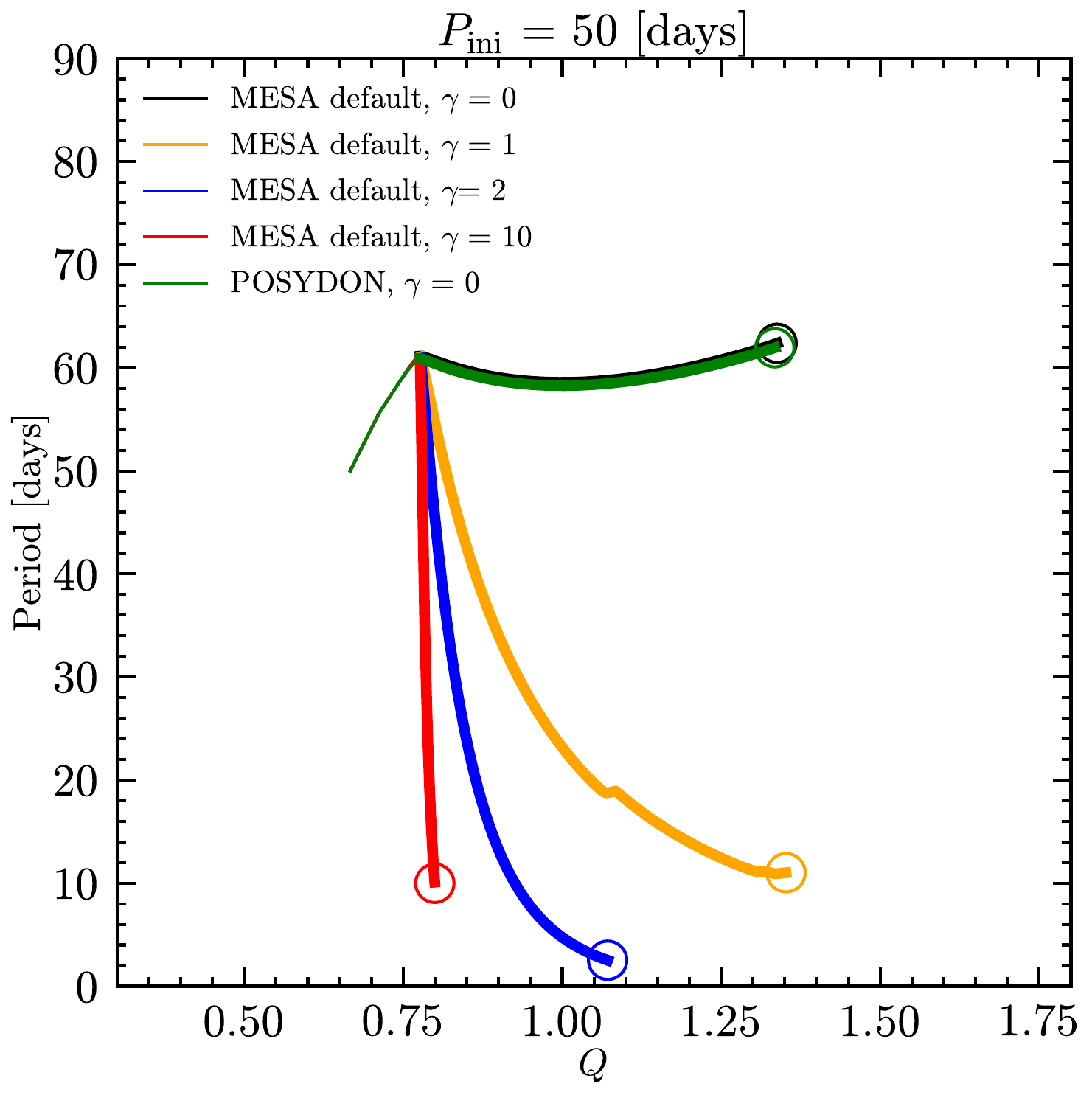}
  \includegraphics[width=0.30\textwidth]{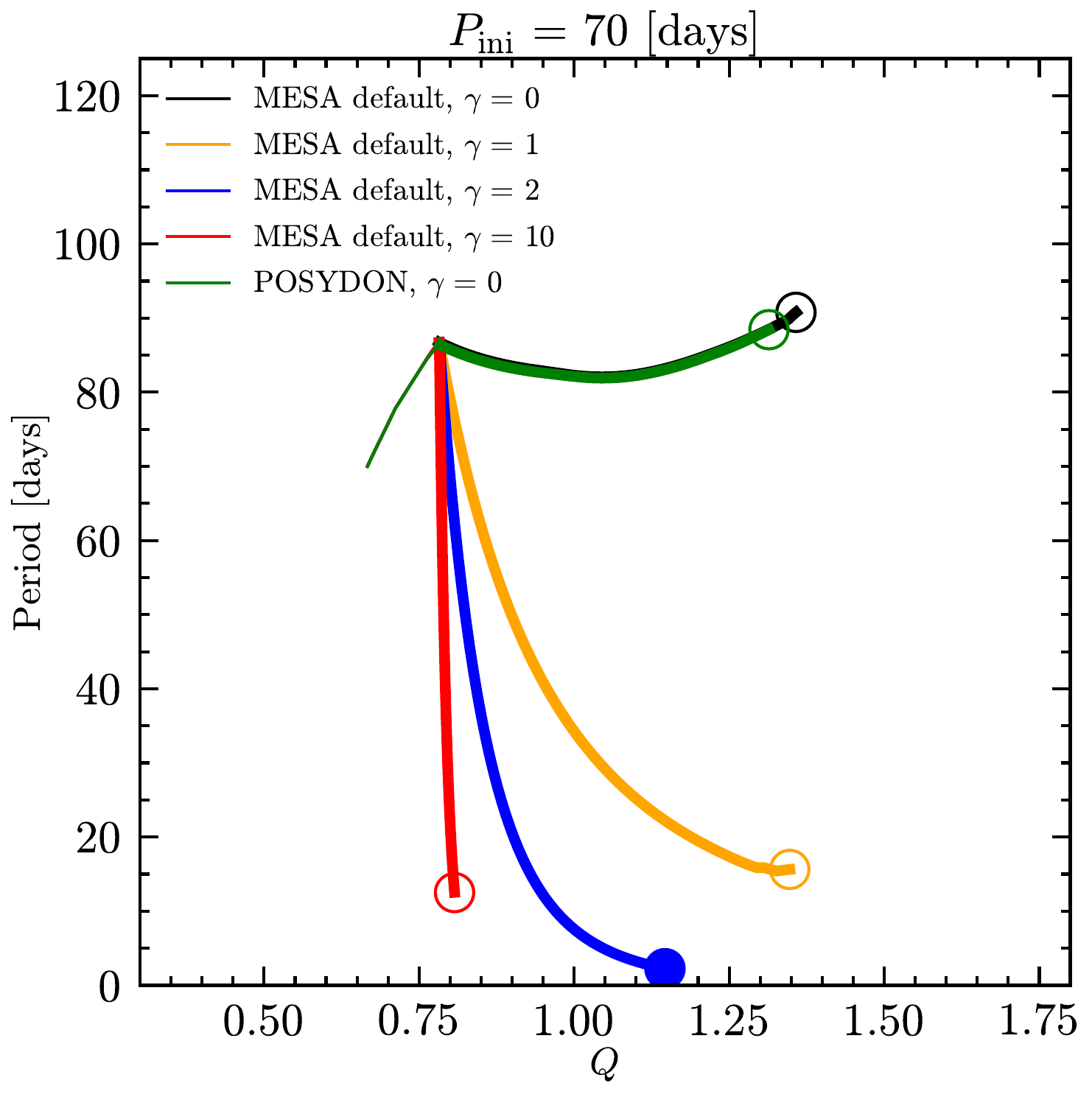}
  \caption{Orbital period vs. mass ratio relation for six cases
    of initial orbital periods (1.25, 3, 5, 10, 50, and 70 days) and masses $M_\mathrm{don,ini}$ = 30 $M_{\odot}$ and $M_\mathrm{acc,ini}$  = 20
    $M_{\odot}$. Different tidal synchronization prescription regimes i.e. \texttt{sync$\_$type="Hut$\_$rad"} (MESA default)
    and \texttt{sync$\_$type="structure$\_$dependent"} (POSYDON) are presented. In addition, different fractions of mass lost from circumbinary toroid (via changing the mass transfer efficiency parameters \texttt{$\delta$} and \texttt{$\gamma$}, see text for details) are shown. The cases with \texttt{$\gamma$} = 0 refer to the conservative mass transfer scenario. The RLOF
    phases are marked by a bold line. Filled and open circles indicate the reasons of track termination: both stars filled their Roche lobes or numerical convergence problem, respectively. In the case of 1.25 day configuration, both stars filled their RL almost simultaneously and we consider that all tracks indicate a subsequent merging.}
  \label{fig:P_q_post}
\end{figure*}

\Figref{fig:P_q_post} illustrates the evolution of orbital period as a
function of mass ratio for six cases of initial period by taking into
account MESA default and POSYDON structure-dependent tide
prescriptions (see \Secref{sec:res_primordial}). In these models,
we wanted to reproduce the post-RLOF phase that would give the
observed period, masses, and mass ratio. With initial periods
above 10 days,
the orbital periods increase as a result of mass transfer
\citep[e.g.,][]{Renzo_2019runaways}, thus the systems do not match the
observed range of periods.
However, varying assumptions on the
  mass transfer efficiency and the angular momentum losses, this
  general trend can change.
 A natural way to prevent the orbital widening, is to assume mass transfer to be nonconservative and remove significant amount of angular momentum, for
example because of the non-accreted material forms a circumbinary disk
(CBD) which applies torques to the inner binary \citep[see e.g.,][]{Vanbeveren_1982,Shao_2016}.

To test this scenario, we explore different mass transfer efficiencies,
using the parametrized analytic formalism of \citet{Soberman_1997}.
Namely, the mass transfer efficiency is defined as the ratio
of the accretor's mass change to the donor's mass change at each timestep, i.e. $\eta$ = $\Delta M_\mathrm{acc}/\Delta M_\mathrm{don}$.
If the mass transfer is nonconservative ($\eta$ $\neq$ 1), the fraction of mass
that is lost to a (coplanar, unmodeled) CBD is defined as $\delta$ = 1 - $\eta$.
Thus, in our simulations, we are continuously updating $\delta$, by assuming three different sizes of the CBD.
Namely, we vary the radius of CBD defined as $r_\mathrm{CBD} = \gamma^2 a$, where $\gamma$ = 1, 2, 10 is another free parameter.
We assume no mass is lost in the vicinity of the accretor and donor ($\alpha$ and $\beta$ equal to zero according to \citealt{Soberman_1997} formalism).

\begin{figure}[!hbpt]
  \centering
  \script{plot_save_mesa_eta.py}
  \includegraphics[width=0.49\textwidth]{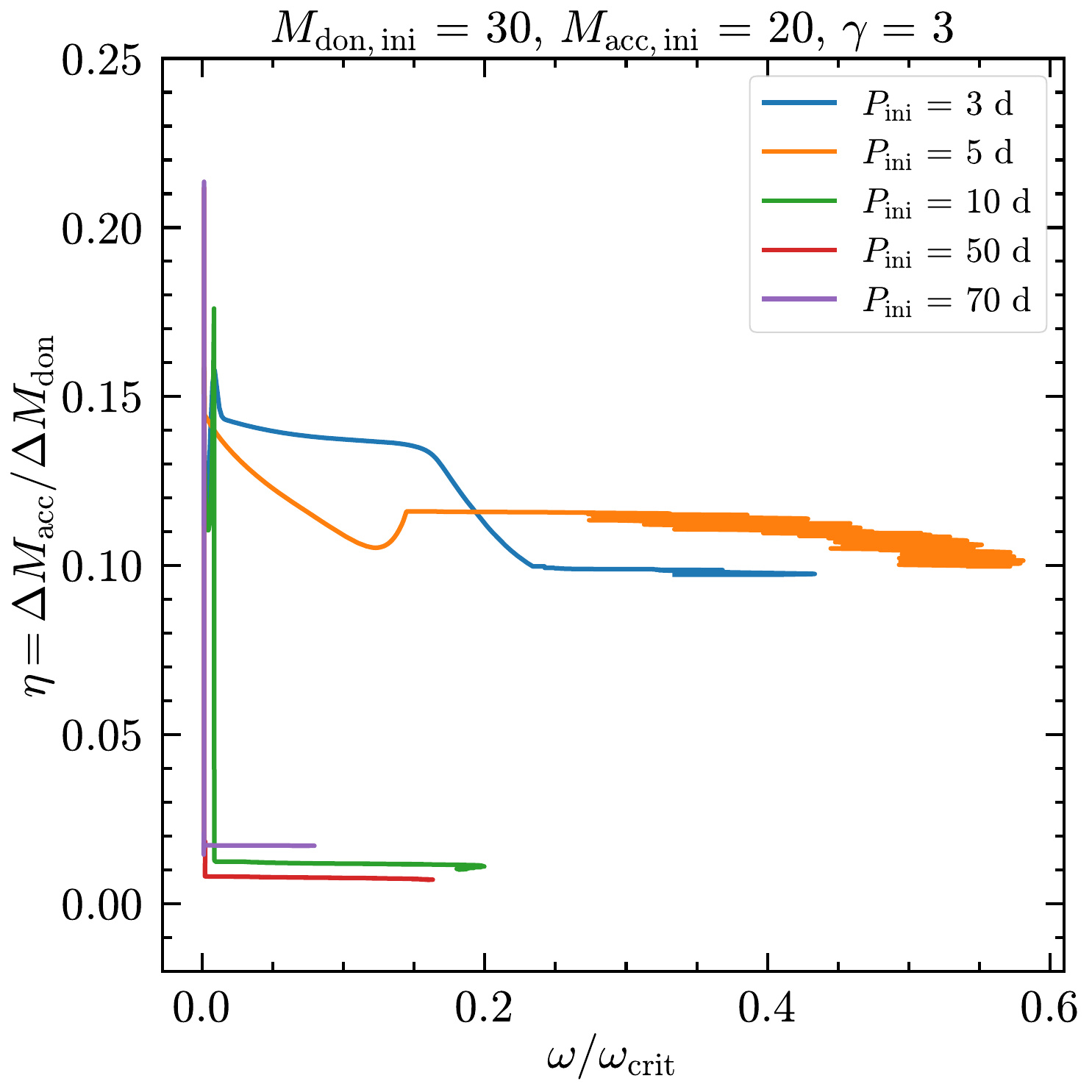}
  \caption{Mass transfer efficiency ($\eta$) as a function of the surface average angular velocity for the various initial periods of a given system.}
  \label{fig:eta_omega}
\end{figure}

We illustrate how the modeled mass transfer efficiency varies as a function of average angular surface velocity in the \Figref{fig:eta_omega} by assuming $\gamma$ = 3 (see the resolution test in \Figref{fig:eta_omega_res}).
As we can see for the short-period models ($P_\mathrm{ini}$ = 3, 5 days), $\eta$ is significantly larger than for the longer period configurations \citep[as it has been also shown by][]{Sen_2022}, which means that mass transfer is not efficient in a such regime of CBD for longer period systems.
It can be explained: with larger orbital angular momentum (longer periods), more material from the donor remains in the CBD, thus the efficiency of accretion becomes low (as can be seen in \Figref{fig:eta_omega}).
Notably, in such a regime of the nonconservative mass transfer, the accretor never reaches the 0.95$\omega_\mathrm{crit}$ threshold --  because for $P$ < 5 days all massive binaries with $Q \sim$ 0.65 reach tidal synchronization early in the MS.
This analysis shows how sensitive is the mass transfer efficiency even in such a simple model of CBD existence.
In addition, we tested binary models with the \citet{fuller:19} AM transport and found this to have a negligible impact on $\eta$ and on the accretor's total angular momentum.
Preliminary analysis shows that with a more efficient AM transport, it takes longer to spin up the accretor with respect to the less efficient AM transport (e.g. considering the Spruit - Tayler dynamo).
To trace the exact history of the mass transfer and occurrence of the CBD we need to explore more complex models of mass transfer which is beyond the scope of the present work.

The behavior of $P$ vs. $Q$ tracks depends on the different assumed values of $\gamma$ and this is presented in \Figref{fig:P_q_post}.
As we can see, for initial periods near the current values ($\sim$ 5 days), the orbital period either increases or decreases, depending on which mass transfer regime we are adopting.
However, the value of $Q$ always remains below 2, i.e., it does not come close to the extreme values that are observed in our systems: as we consider the post-interaction paradigm, the modeled $Q$ should be the inverse of the observed ($q$) i.e. our systems should have $Q$ > 5.

For shorter initial periods, the period further decreases, taking us away from the observed values.
Moreover, in the majority of cases, the systems merge with the most extreme case ($P_\mathrm{ini}$ = 1.25 days) within 500 -- 800 years following the onset of RLOF (see the first top panel of
\Figref{fig:P_q_post}). Thus, during and after the donor's RLOF stage, we can not reproduce the observed periods and mass ratios whatever the value of $P_\mathrm{ini}$.
Further simulations show, that in the majority of cases, the systems with periods less than 10 days will also merge. We identified the time of merging events as the time at which both the secondaries and the primaries fill their Roche Lobe and we indicated this by filled circles at the end of each track in \Figref{fig:P_q_post}.

In the case of larger initial periods ($P_\mathrm{ini}$ $>$ 10 days, three bottom panels in \Figref{fig:P_q_post}), nonconservative mass transfer can form short-period systems, however, extreme mass ratios would be again difficult to reach.
To confirm these results, we compute an additional set of models with different initial masses of the components.
The results of these simulations confirm the main increasing trend of the mass ratio up to the possible merging of the systems (in a nonconservative mass transfer regime, in terms of $\eta$ > 0) or up to the moment of system detachment (at conservative mass transfer regime, when  $\eta$ = 0).
However, the value reached remain far from the observed ones.
We can therefore conclude that the presented scenario to form short-period, extreme mass ratio, fast-rotating binaries from binary interaction is not favored.

Short-period binaries could form through a
post CE scenario \citep[see channel 4 in][]{Willems_2004}.
However, if we assume a post-CE origin of our systems, the rotational velocity of the current primary component will be conditioned only by tidal synchronization. As demonstrated in \Secref{sec:pre-interaction}, it is not possible to reach the observed \vsini~ (exceeding tidal synchronization values), especially in the case of HD\,46485.
A post-common envelope ejection origin of our targets thus appears to be unlikely.

\section{Merger in a triple}
\label{sec:triple_scenario}

The majority of massive OB-type stars are born in high-multiplicity
systems, with $\sim60\%$ having at least two companions, i.e. being in triple systems
\citep[e.g.,][]{moe_2017, Offner_2022}. Thus, we also consider a
triple-evolution scenario to explain the rotation of the observed star
in these systems. As we demonstrate below, the present day observed
mass ratios of HD\,25631, HD\,191495, and HD\,46485 imply that if they
had a past as triple stars, they were hierarchical, $m_1+m_2 \gg m_3$,
where $m_i$ are the individual stellar masses (contrary to the previous
sections where lower case $m$ was used for mass-coordinate inside the star).

In this scenario, the current primary star in the observed system is
the remnant of the merger of the inner binary in an originally triple
system. The observed fast rotation comes from the (fraction of)
orbital angular momentum of the inner binary retained by the merger
product \citep[e.g.,][]{deMink_2014}. The observed present-day binary
orbit corresponds to the initial outer binary orbit, possibly modified
by the mass and angular momentum losses during the merger itself.
Here, we discuss the reasons why we consider this scenario unlikely
for HD\,25631, HD\,191495, and HD\,46485, although it is hard to
completely rule it out given the vast ten-dimensional parameter space
for initial triple configurations \citep[e.g.,][]{toonen:16}.
\paragraph{The mergers and fast rotation.}
Some multidimensional magneto-hydrodynamics simulations of mergers between
main-sequence stars of masses comparable to ours ($\sim 9+8\,M_\odot$)
predicted the remnant to be slow rotating and magnetic
(\citealt{schneider:16, schneider:19, schneider:20}, see also
\citealt{wang:20}). The reason for the slow post-merger rotation is
three-fold: (\emph{i}) dynamical mass and angular-momentum loss during
the merger itself, (\emph{ii}) magnetic-braking post-merger, and most
importantly, (\emph{iii}) the internal readjustment of the merger
product. In fact, the merger happens on a dynamical timescale,
therefore the core structure of the initially more massive star is
``frozen'' during the process. The core settles in the center of the
merger product, but initially retains its pre-merger density, set by
the structure of a lower mass star. Post-merger, the structure has to readjust to the new and increased total mass by lowering its
central density. Thus, on a thermal timescale, the core expands
consuming angular momentum \citep{schneider:19, schneider:20}.

In addition, the empirical initial rotation distributions for apparently
single stars \citep{vfts_2013_otype} and confirmed binaries
\citep{vfts_2015_otype} in 30 Doradus suggest that the fastest
rotators detected are binary accretors \citep[e.g.,][]{packet:81,
  blaauw:93, Renzo_2021} and possibly mergers
\citep[e.g.,][]{deMink_2013}. In fact, theoretical predictions for
the surface and internal rotation of merger products can vary
significantly as a function of mass ratio and evolutionary stage of
the pre-merger binary \citep{chatzopoulos:20, renzo:20c}. The
angular momentum transport before, during, and after the merger is also
highly uncertain \citep[e.g., see discussion in][]{spruit:99,
  spruit:02, fuller:19, denhartogh:20}. For example,
\cite{chatzopoulos:20} proposed that Betelgeuse's fast rotation (for a
red-supergiant) may be explained with a post-MS
merger. Because of these uncertainties, we do not consider the theoretical expectation of slow
rotating merger products sufficient to discard a
merger-in-a-triple scenario on its own.

\paragraph{Dynamical instability of putative triple progenitor.}
The present-day observed periods, corresponding to orbital separations of
$\sim24.6\,R_\odot$, $24.7\,R_\odot$, and $45\,R_\odot$, for HD\,25631, HD\,191495, and HD\,46485,
respectively, put strong constraints on the initial conditions for a
triple.

We treat the orbital evolution during the merger by analogy with the
``Blaauw kick'', typically considered in the context of spherically
symmetric\footnote{Meaning no ``natal kick'' is imparted by
  asymmetries in the mass-shedding event, the explosion or, in our
  triple scenario, the merger.} supernova explosions in a binary
\citep{blaauw:61, boersma:61}. The merger ejects a fraction $f_\Delta$
of the total mass of the inner binary. Hydrodynamical simulations of
stellar collisions, such as the ones that may occur in an unstable
triple system, usually produce $f_\Delta\lesssim 12\%$
\citep[e.g.,][]{lombardi:02, glebbeek:13, renzo:20c, ballone:23}, so
we explore two values of $f_\Delta=0$ and $0.1$. We assume the ejecta
leave the (triple) system instantaneously without interacting with the
outer companion. \Secref{sec:caveats_triple} discusses further these
approximations.

Throughout this section, we neglect the orbital widening caused by
wind mass loss. In fact, the most massive star considered here is
$\sim24\,M_\odot$ and has an apparent age of $\sim2$\,Myr.
Theoretically, a solar-metallicity star of comparable mass loses
$\lesssim 4-8\%$ of its total mass over the entire main-sequence
lasting $\sim 7$\,Myr \citep{renzo:17}. Thus, we can assume that the
present-day observed primary mass $M_1$ is the post-merger mass within
a few percent precision. Correcting for the fraction of mass
lost at merger $f_\Delta$, gives the pre-merger total mass of the inner
binary:
\begin{equation}
  \label{eq:mtot_in}
  M_1 = m_1+m_2 - f_\Delta(m_1+m_2) \Rightarrow m_1+m_2 = \frac{M_1}{1-f_\Delta} \ \ ,
\end{equation}
and since by construction $m_1$ and $m_2$ are individually smaller
than $M_1$, we can again neglect wind mass loss rates pre-merger and
consider $m_1+m_2$ the total ZAMS mass of the
inner binary. Similarly, neglecting wind mass loss, at ZAMS the
tertiary star has a very similar mass as the present day secondary mass
$m_3=M_2$. Thus the outer binary mass ratio is known
$q_\mathrm{out}=m_3/(m_1+m_2)= M_2(1-f_\Delta)/M_1$.

\begin{figure}[hbtp]
  \centering
  \script{plot_min_a_in.py}
  \includegraphics[width=0.5\textwidth]{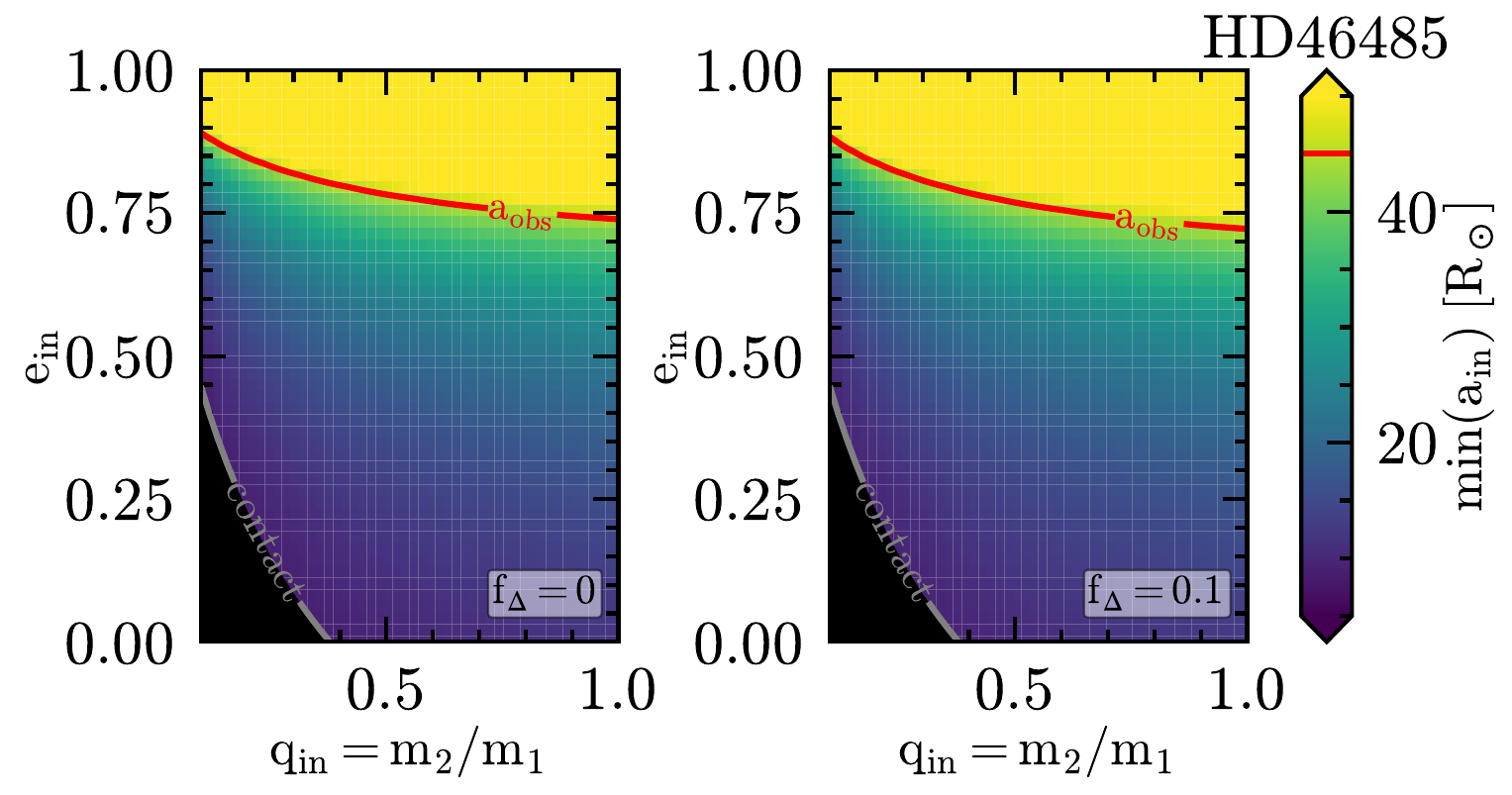}
  \includegraphics[width=0.5\textwidth]{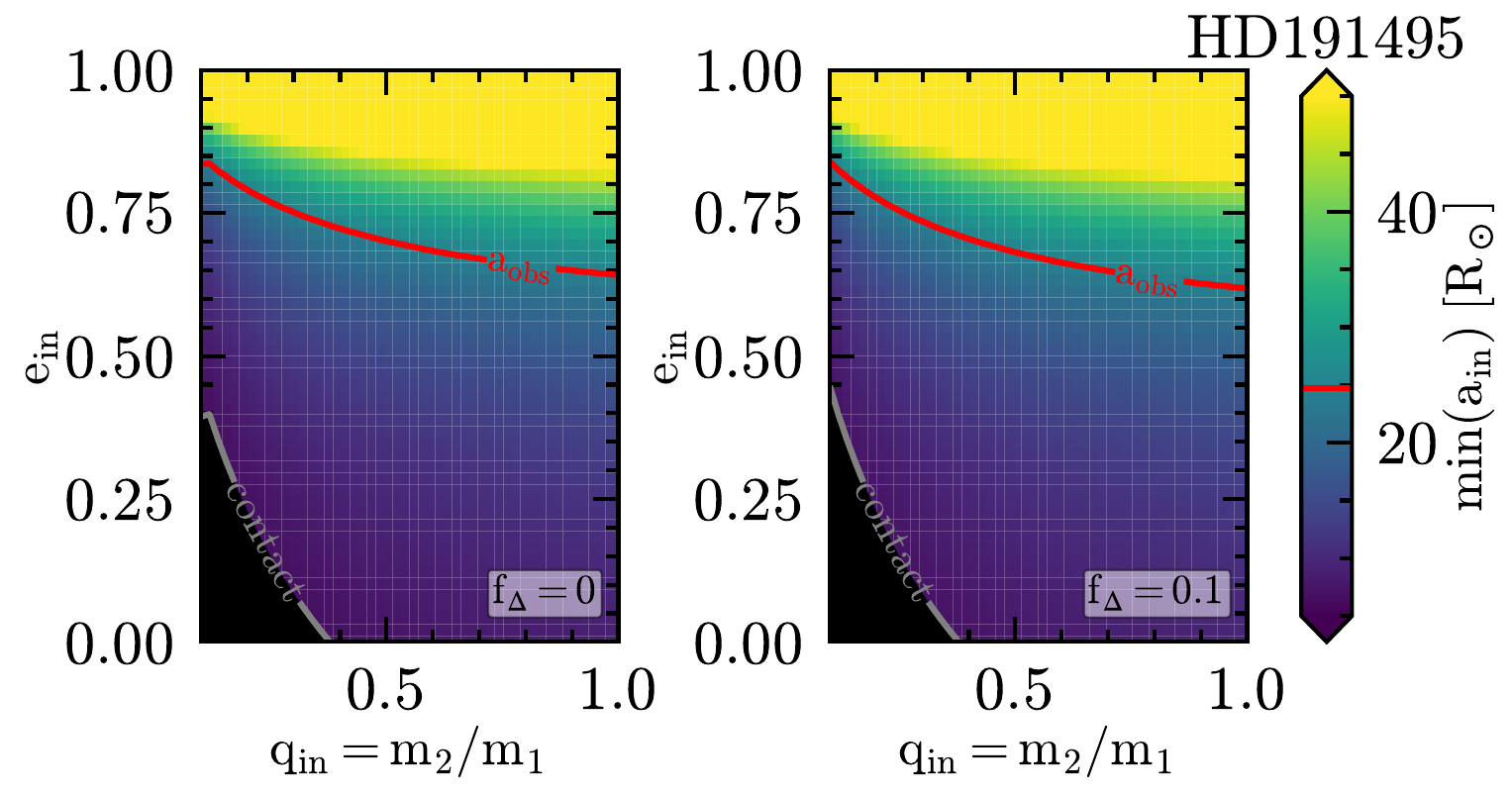}
  \includegraphics[width=0.5\textwidth]{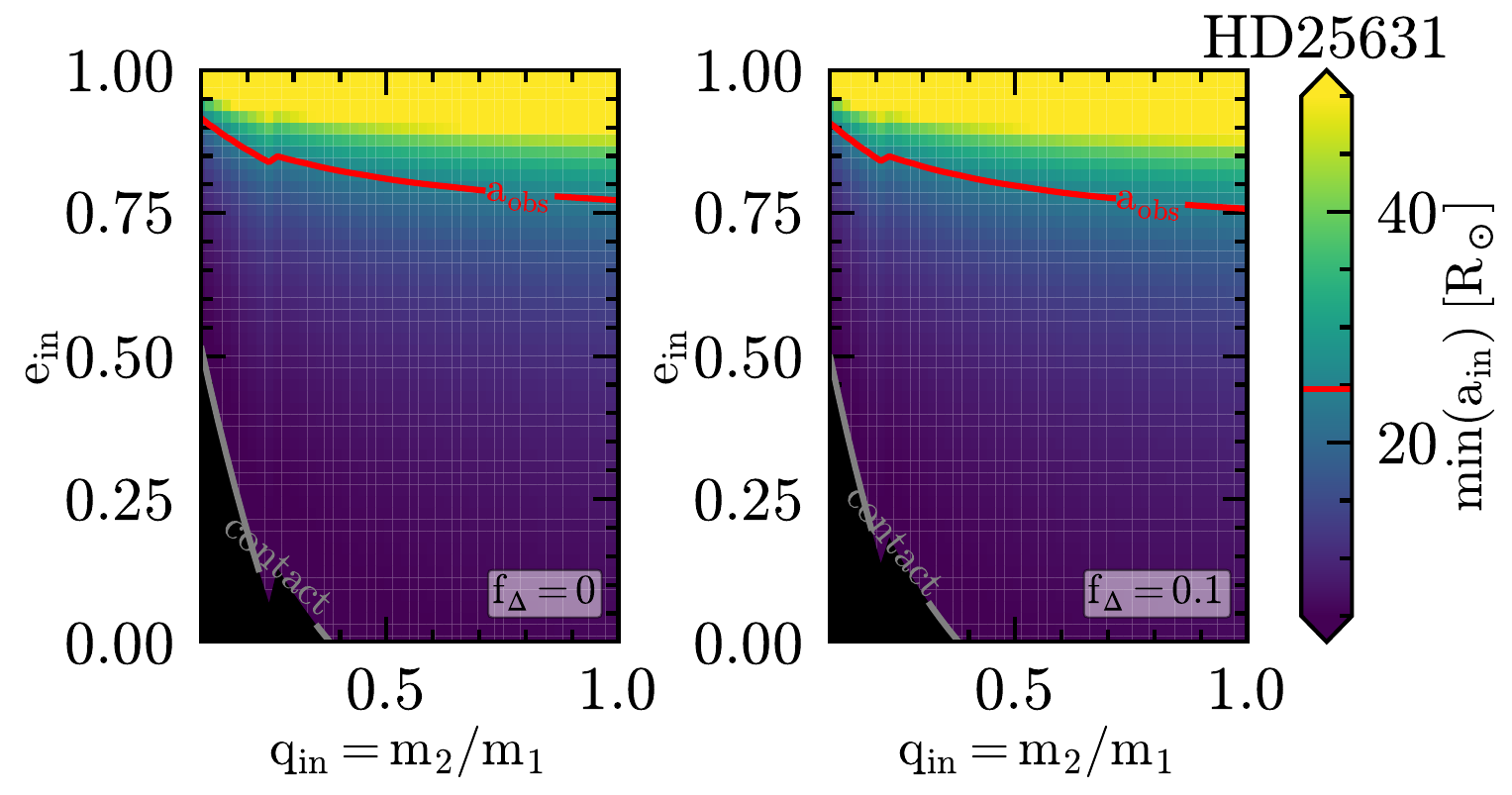}
  \caption{Minimum initial inner binary separation
    $\min(a_\mathrm{in})$ to prevent RLOF at periastron at ZAMS (i.e., $R_1(m_1) \leq R_\mathrm{RL,1}$ and
    $R_2(m_2)\leq R_\mathrm{RL,2}$) as a function of the inner binary
    mass ratio $q_\mathrm{in}=m_2/m_1$ and eccentricity
    $e_\mathrm{in}$ for HD45485 (top), HD191495 (middle), and HD25631
    (bottom). The red solid lines mark the present-day observed
    separation, corresponding to the evolved post-merger
    $a_\mathrm{out}$ in the merger-in-a-triple scenario, while the
    black area in the bottom left corners corresponds the unphysical
    region with initial inner binary separation with $R_i(m_i) \leq R_{\mathrm{RL},i}$. In the bottom row, the
    boundary of this unphysical region has a kink caused by the change
    of slope in the assumed mass-radius relation: HD\,25631 primary
    mass is $M_{1}\simeq 7\,M_\odot$, so for
    $q_\mathrm{in}\lesssim 0.2$ (the exact threshold depends on
    $f_\Delta$), the expected values of $m_2$ are sufficiently low
    that the exponent $\xi$ in the
    $R_\mathrm{ZAMS}\equiv R_\mathrm{ZAMS}(m)$ changes. For the right
    column with $f_\Delta\neq0$, our assumptions based on the
    \cite{blaauw:61} approximation may be inconsistent (see \Secref{sec:caveats_triple}).}
  \label{fig:min_a_in}
\end{figure}

We can then find the range of initial inner separations
$a_\mathrm{in}$ possible as a function of the unknown eccentricity
$e_\mathrm{in}$ and mass ratio $q_\mathrm{in}$ of the inner binary,
shown in \Figref{fig:min_a_in}. We define a lower boundary
$\min(a_\mathrm{in})$ corresponding to inner binaries where both stars
are filling their Roche lobe at periastron at ZAMS. To calculate the
Roche lobe size at ZAMS, we use the \cite{eggleton:83} fitting formula
$R_{\mathrm{RL},i} = a E(q_i) \equiv R_{\mathrm{RL},i}(q_i)$ where $a$
is the binary separation $E(q_i)$ is an analytic function of the mass
ratio $q_1=m_1/m_2$ and $q_2=m_2/m_1$. Although the Roche geometry is
valid only in the restricted \emph{circular} two-body problem, we approximately
account for the eccentricity of the inner binary by approximating the
elliptical orbit with a circular orbit with radius equal to the
periastron, substituting
$a_\mathrm{in}\rightarrow a_\mathrm{in}(1-e_\mathrm{in})$. Thus, for
each ($q_\mathrm{in}$, $e_\mathrm{in}$) pair we can impose the condition
$R_{\mathrm{ZAMS},i} \leq R_{\mathrm{RL}, i}\equiv R_{\mathrm{RL}, i}(q_\mathrm{in}, e_\mathrm{in})$
and translate
it into
$a\geq \min(a_\mathrm{in}) \equiv \min(a_\mathrm{in})(q_\mathrm{in},e_\mathrm{in})$.
For simplicity, we estimate the initial radii $R_{\mathrm{ZAMS}, i}$
using an analytic mass-radius relation for homogeneous polytropic
stars, $R_{\mathrm{ZAMS} ,i}= R_\odot(m_i/M_\odot)^{\xi}$ with
$\xi=0.9$ for $M_i<1.4\,M_\odot$ and $\xi=0.6$ otherwise. Note that
introducing a constraint depending on the ZAMS radii introduces a
mass-dependence that breaks the scale-free nature of the Newtonian
three-body problem.

With the assumption of instantaneous mass loss from the inner binary
\citep{blaauw:61}, the outer binary is either unperturbed
($f_\Delta=0$) or it widens and increases its eccentricity
($f_\Delta>0$). Thus, we expect the outer binary to have an initial
separation shorter than that of the binaries observed today:
\begin{equation}
  \label{eq:sep_ordering}
  a_\mathrm{obs} \gtrsim a_\mathrm{out,post-merger}\gtrsim
  a_\mathrm{out, pre-merger}\simeq a_\mathrm{out, ZAMS} \ \,
\end{equation}
where we use the results of \Secref{sec:pre-interaction} to assume the binary is presently widening, once-again, we neglect widening caused by wind mass loss and
secular orbital effects, and by definition
$a_\mathrm{out,ZAMS} \gtrsim a_\mathrm{in, ZAMS} \gtrsim \min(a_\mathrm{in})$
(see \Figref{fig:min_a_in}). The \cite{blaauw:61} model assumes
circular pre-mass-ejection binaries. Given the small observed
eccentricities and the relative inefficiency of tides discussed in
\Secref{sec:pre-interaction}, we consider here pre-merger outer
eccentricity equal to today's observed value
$e_\mathrm{out}=e_\mathrm{obs}\simeq0$, which is consistent with
$f_\Delta=0$ but in possible tension with nonzero mass loss mergers
$f_\Delta\neq0$ (see \Secref{sec:caveats_triple}).

The triple merger scenario is meaningfully different from a pure
binary evolution model only if the triple is initially dynamically stable and gets
destabilized by stellar evolution processes (e.g., radial expansion,
mass loss, tides, \citealt{perets:12}). If that is not the case, the
triple will either evolve chaotically (leading to a merger or
disruption within a few Lyapunov times) or is unstable (leading to
merger or disruption within a few dynamical times set by the inner/outer
orbital period). Either of these scenarios would result in the prompt
formation of a binary (compared to stellar evolution timescales) and
reduce to the pre-interaction scenario discussed in
\Secref{sec:pre-interaction}. Conversely if a stable triple can form,
the mass hierarchy implied by the present-day binaries suggests
$m_1+m_2 \gg m_3$ for most measurable values of
$q_\mathrm{in}=m_2/m_1$. Therefore, in the original inner binary
stellar evolution proceeds much faster, and instabilities in the
triple triggered by the stellar evolution in the inner binary are
likely \citep{perets:12}, so we can limit our discussion to initially
 stable and non-chaotic triples.

 With the physical constraints discussed above, we can build triple
 systems compatible with the observed binaries and
 quantify their dynamical stability. To explore the
 parameter space for triple systems, we draw $10^6$ initial triples for each
 observed binary and
 evaluate their dynamical stability\footnote{The resulting triple
   populations with an estimate of their stability are
 publicly available at \url{https://doi.org/10.5281/zenodo.10028333}}.
 We sample a uniform distributions in the range $0\leq e_\mathrm{in}<1$ for the
 inner eccentricity, $0.1\leq q_\mathrm{in}\leq 1$ for the inner
 binary mass ratio,
 $\min(a_\mathrm{in})\leq a_\mathrm{in}\leq a_\mathrm{out}\leq a_\mathrm{obs}$
 for the inner and outer separations where $a_\mathrm{obs}$ is the
 present-day observed binary separation,
 $0\leq i_\mathrm{rel}\leq \pi$ for the mutual inclination of the
 inner and outer orbit, and $0\leq e_\mathrm{out}\leq e_\mathrm{obs}$
 for the outer eccentricity. Certain choice of $f_\Delta$,
 $q_\mathrm{in}$, and $e_\mathrm{in}$ result in the unphysical
 condition
 $\min(a_\mathrm{in}) \geq \max(a_\mathrm{out})=a_\mathrm{obs}$. This
 means that no triple can form with the assumed parameters within the
 constraints of the scenario: we discard these systems leaving us with
 $\sim 8\times 10^5$ triples for each $f_\Delta$ and each observed
 binary.

 Recently, machine-learning techniques have been applied to the
 determination of the dynamical stability of a triple system
 \citep[e.g.,][]{lalande:22, vynatheya:22, vynatheya:23}. Here, we use the
 publicly available ``ghost orbit'' method presented in
 \cite{vynatheya:23}, which for every triple defined by the set of
 parameters
 $(q_\mathrm{in},\ q_\mathrm{out},\ a_\mathrm{in}/a_\mathrm{out},\ e_\mathrm{in},\ e_\mathrm{out},\ i_\mathrm{rel})$
 provides a probability for the system to be dynamically unstable
 and/or chaotic $P$ as a function of the initial architecture of the
 triple.

 The stability criterion is based on the divergence of orbits initially
 differing slightly in $a_\mathrm{in}$ in direct N-body simulations
 over a time corresponding to $\sim 100$ outer-periods. We emphasize
 that this does not account for close stellar encounters resulting
 in mergers during a chaotic phase of evolution because the stars are
 modeled as point masses, therefore $P$ is really a measure of the
 probability of the triple system to be dynamically disrupted within
 100 periods of the outer binary. The underlying N-body simulation
 neglects the arguments of periastron parameters (in other words, the
 relative phase of the inner and outer binary), which is known to
 affect the stability, and do not include general relativistic effects
 (not relevant for stellar triples). The performance of the model is
 slightly worse on retrograde orbit, but nevertheless improves over
 the semi-analytic stability criterion from \cite{mardling:99} and
 \cite{vynatheya:22}.

\begin{figure}[htbp]
     \script{plot_P_unstable.py}
   \includegraphics[width=0.5\textwidth]{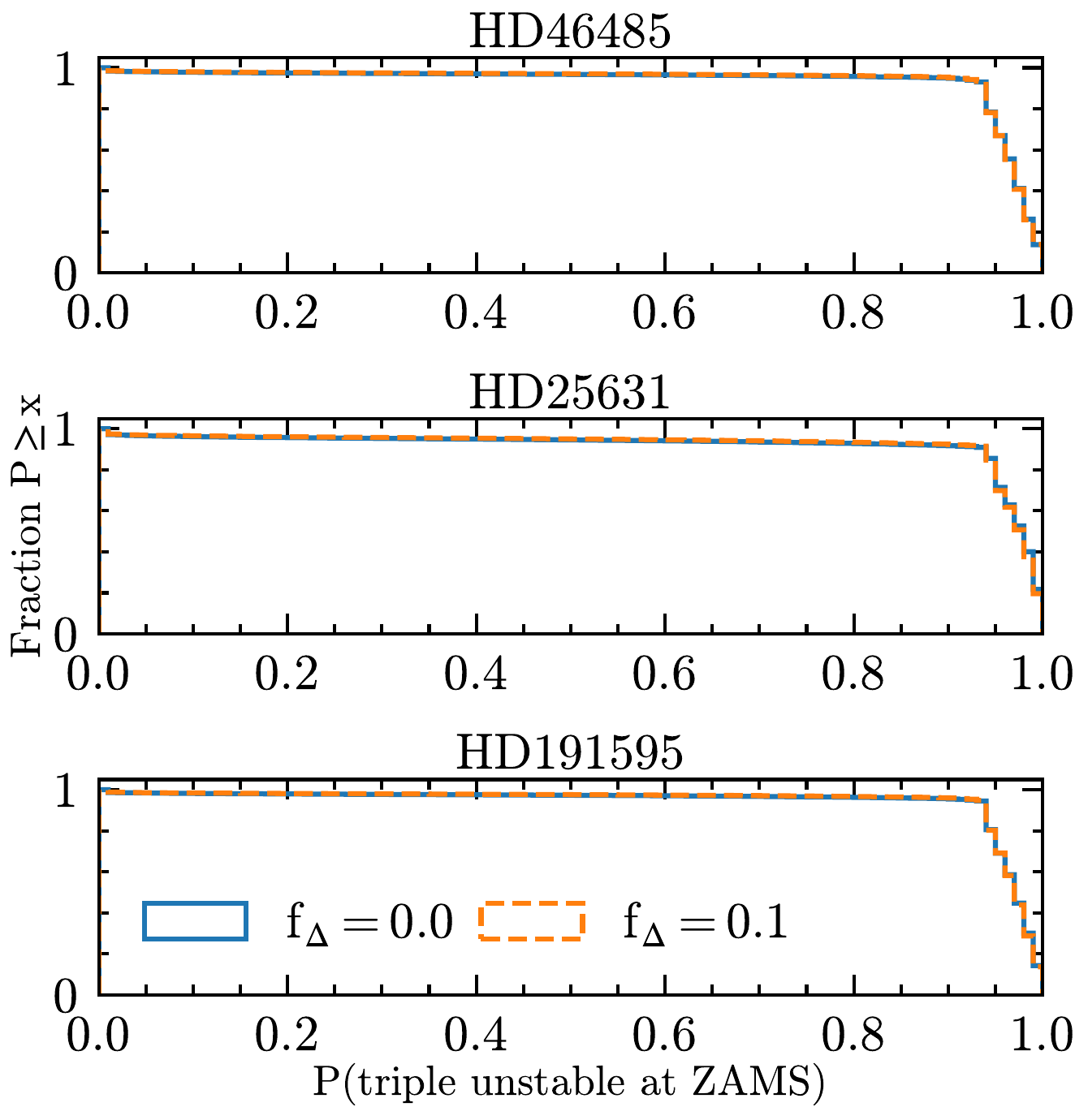}
   \caption{Vast majority of triple systems compatible with the
     ``merger in a triple'' scenario has a probability of being
     dynamically unstable $\gtrsim90\%$. For each observed binary, we
     show the fraction of compatible putative triple progenitors that
     has a probability of being disrupted larger than the x coordinate
     according to the ``ghost orbit'' method of \cite{vynatheya:23}.
     Orange and blue curves correspond to the two values of fractional
     mass loss at merger $f_\Delta=0$ and 1 assumed. The distribution
     is marginalized over all parameters defining the architecture of
     the triples compatible with the present-day binaries (see text).}
   \label{fig:ML_stability}
 \end{figure}

 Each row in \Figref{fig:ML_stability} shows the complementary
 cumulative distribution of the sampled triples as a function of the
 probability of being unstable for each of the observed binaries. For
 each curve, the y-value represents the fraction of triple systems
 that have probability $P$ of being unstable greater or equal than the
 abscissa value, marginalized over all parameters. Regardless of the
 observed binary considered, $\gtrsim 95\%$ of possible triple systems
 have a probability $\gtrsim 90\%$ of being dynamically unstable,
 meaning they would disrupt within $\lesssim100$ outer orbital periods.
 The longest period binary has
 $7\,\mathrm{days}\simeq P_\mathrm{obs}\gtrsim P_\mathrm{out, pre-merger}$
 (cf.~\Eqref{eq:sep_ordering}), this means the timescale for
 disruption is
 $\lesssim 100\times P_\mathrm{obs} \lesssim 700\,\mathrm{days}\simeq 2\,\mathrm{years}$.
 Since the initial time is rather arbitrarily chosen to be at ZAMS
 (which is not physically defined with such temporal precision), we
 argue that such short times make the ``triple merger scenario''
 virtually indistinguishable from the ``primordial rotation
 scenario'': the merger has to occur before dynamical disruption of
 the system, therefore so close to ZAMS that it can be considered as
 part of the star formation process. We note also that the probability
 $P$ is loosely correlated with the number of outer orbits completed
 before disruption, and the high values shown in
 \Figref{fig:ML_stability} suggest the vast majority of the triples
 would disrupt in much less than 100 outer orbits, shortening the time
 since ZAMS even further.

 Nevertheless, the scenario is not completely ruled out, since a few
 percent of the simulated triple systems result in a very low
 probability of being unstable.
Refining our understanding of the future evolution of survival initial triples, which depends on many initial parameters, would require an additional effort in numerical modeling \citep[e.g.][]{toonen:20,Dorozsmai_2024}, which is beyond the scope of present work.
Nevertheless, we note that most past triples are expected to be eccentric, which is not the case in our systems.

 \subsection{Caveats}
\label{sec:caveats_triple}
Our arguments above require a few words of caution. The stability
criterion we employ (see \citealt{vynatheya:23}) ignores the
occurrence of stellar mergers in the triple system, and is in this
sense conservatively underestimating the probability of dynamical
instability (a triple may be unstable because of mergers rather than
disruption).

More importantly, if $f_\Delta\neq0$, we have assumed that the merger
ejecta leave the outer binary instantaneously and without
interacting with the third companion. These assumptions result in an
instantaneous widening and an increase in eccentricity of the outer
binary. For the
original application of \cite{blaauw:61} to supernova ejecta, this
assumption is justified since ejecta velocity are
$\sim10^4\,\mathrm{km\ s^{-1}} \gg v_\mathrm{orb}$ where
$v_\mathrm{orb}\simeq10-100\,\mathrm{km\ s^{-1}}$ is the orbital
velocity of the binary containing the explosion. For our case of
merger in a binary, with the constraints of the outer binary having a
smaller orbital separation than today's observed separation,
this approximation is not as good and can even formally lead to
absurd conclusion. We can estimate the ejecta velocity from the escape velocity
of the inner binary
$v_\mathrm{ejecta}\simeq \sqrt{2 G (m_1+m_2)/a_\mathrm{in}}$ and use
Kepler's third law to obtain the outer orbital velocity
$v_\mathrm{orb, out}=\sqrt{G(m_1+m_2+m_3)/a_\mathrm{out}}$. The
condition $v_\mathrm{ejecta}\gg v_\mathrm{orb, out}$ reduces to
$a_\mathrm{out}/a_\mathrm{in} \gg 2(m_1+m_2)/(m_1+m_2+m_3)$ which may
not be possible to satisfy (depending on $f_\Delta$). In most cases,
our triple systems would have
$v_\mathrm{ejecta}\simeq v_\mathrm{orb, out}$, which would require
detailed (radiation)-hydrodynamical simulations of the interaction between
the merger ejecta with the outer orbit to improve on the crude
approximation of the \cite{blaauw:61} model. If the ejecta remain
sufficiently dense as they cross $a_\mathrm{out}$ they may lead to a
``triple common envelope'' \citep[e.g.,][]{toonen:20} and
$a_\mathrm{out}$ may shrink rather than expand because of the merger.
Nevertheless, as discussed in the context of \Figref{fig:v_surf_acceleration}, post-common envelope tidal spin-up is not a valuable scenario for the mass ratio and ages observed.

\section{Discussion}
\label{sec:discussions}

Based on our results described in the previous section, we conclude
  that primary stars in HD\,25631, HD\,191495, and HD\,46485 are spinning faster than tidal synchronization most
probably because of the results of star formation processes (pre-interaction scenario, \Secref{sec:pre-interaction}) rather than binary
interactions (\Secref{sec:post-interaction}) or a merger in a triple system (\Secref{sec:triple_scenario}).

\begin{table*}[!htbp]
{\small
\caption{Basic physical and orbital parameters for systems with a known fast-rotating OB star, extreme mass ratio, and short periods. The list of stars is ordered by decreasing mass ratio estimates.}
\label{table:ref_fast}
\begin{tabular}{l |c c c c c c c}
\hline
\hline
Target                  & SpC    &  \vsini  & $P$ & $q$ &
                                                                       $e$  & Companion  & Reference \\
& & [\kms] & [d] & & & & \\
\hline
HD\,163892        &    O9.5 IV(n)       &     201       &    7.8     &  0.17              &   0.04     &  B5-7 star     &   \citet{Mahy_2022}     \\
HD\,165246        &    O8 V(n)          &     254         &    4.6     &  0.16               &   0.03   &  pre-MS star  &   \citet{Johnston2021}       \\
HD\,152200        &   O9.7 II-IIIn       &     205        &    4.5     &     0.12             &  0.17     &  unknown $^{\ast}$   &   \citet{Britavskiy_2023}     \\
HD\,46485         &   O7 V ((f))nvar?    &     334        &    6.9     &     0.035 -- 0.045      &  0     &  pre-MS star     &   \citet{Naze_2023_rot}     \\
HD\,138690 ($\gamma$ Lup A)  &   B2Vn     &     236        &    2.8     &     0.1 -- 0.234            &  0.10    &  pre-MS star     &   \citet{Jerzykiewicz_2021}     \\
HD\,25631         &   B3 V    &     221        &    5.2     &     0.12 -- 0.15             &  0     &  pre-MS star     &   \citet{Naze_2023_rot}     \\
HD\,191495         &   B0 V    &     201        &    3.6     &     0.08 -- 0.12             &  0     &  pre-MS star     &   \citet{Naze_2023_rot}     \\
HD\,149834             &   B type     &     216        &    4.9     &     0.09             &  0.04     &  pre-MS star     &   \citet{Stassun_2021}   \\
HD\,58730      &   late-B type   &     183 $^{\ast\ast}$       &    3.6     &     0.07            &  $<$ 0.07     &  pre-MS M type star     &   \citet{Stevens_2020}     \\
\hline
\end{tabular}
\\
{\bf Notes.}  $^{\ast}$ The nature of the secondary component in HD\,152200 system requires further investigateion. $^{\ast\ast}$ This system does not pass our \vsini~ selection criteria. }
\end{table*}
\subsection{Observational context}

We list in \Tabref{table:ref_fast} all systems (including our targets) for which both
  photometric and spectroscopic analyses are available and the derived physical parameters are similar to our
systems. Namely, we look for short-period OB-type massive binaries
($P\lesssim 10$\,days) with an extreme mass ratio ($q$ $\lesssim$ 0.2), and
fast rotation (\vsini~ $\gtrsim$ 200\,\kms).

In the majority of cases, the resulting orbital solutions and spectroscopic disentangling suggest that
  identified O-type eclipsing binaries typically have pre-MS secondary
  components, which is in line with the detached pre-interaction phase being
  astrophysically long w.r.t. stellar lifetimes (i.e., HD\,165246). Interestingly,
according to \citet{Mahy_2022}, more long-period and highly eccentric
binaries among the initial sample of SB1 fast rotators from
\citet{Britavskiy_2023} could possibly host a neutron
star. However, according to their spectroscopic disentangling simulations, only
the secondaries with  $M_2\gtrsim 3\,M_{\odot}$ can be detected in the
available spectra. Thus, there is a possibility that the secondary components are
  low-mass stars in pre-MS or MS stage \citep[e.g., HD\,15137 and HD\,165174 in][]{Mahy_2022}.
That is in agreement with the $P$ vs. $q$ behavior shown in \Figref{fig:P_q_post} for conservative mass
transfer in \Secref{sec:post-interaction}: it is more likely to observe a post-interacting system in long period configurations.

Regarding B-type stars, in addition to the ones investigated in the present work, there are four other short-period binaries known with an extreme mass ratio, namely $\nu$ Cen, $\gamma$ Lup A \citep[see,][]{Jerzykiewicz_2021}, HD\,58730 \citep{Stevens_2020}, and HD\,149834 \citep{Stassun_2021}.
Companions in all these binaries are pre-MS stars, interestingly that HD\,149834 and $\gamma$ Lup A,  harbour fast-rotating B-type stars (\vsini~= 216 $\pm$ 11 \kms, \vsini~= 236 $\pm$ 5 \kms~ respectively), while $\nu$ Cen is a slow rotator with a \vsini~=  65 $\pm$ 6 \kms.
Also, HD\,58730 is just below our rotation limit i.e. \vsini~= 183 $\pm$ 1 \kms.
In addition, about 20 short-period binaries with an extreme mass ratio have been detected in the Large Magellanic Cloud \citep[see][]{Moe_2015}. The lightcurves of these binaries are very similar to the objects we investigate in the present work, and \citeauthor{Moe_2015} suggest a pre-MS nature for the majority of the companions.
Unfortunately, no spectroscopic analysis was performed so there is no information regarding the rotation rate of this sample of stars nor any radial velocity curves for these systems.
In this context, there are also a couple of fast-rotating SB1 systems known in the Large Magellanic Cloud \citep[O-type primaries, see][]{vfts_2015_otype} and in the Small Magellanic Cloud \citep[B-type primaries, see][]{Bodensteiner_2023}. However, the orbital parameters of these binaries are yet unknown.

As we can see, all the stars from \Tabref{table:ref_fast} with
orbital periods $P$ $<$ 10 days have pre-MS or main-sequence companions in a pre-interacting stage.
We can then draw the natural conclusion that the fast rotation of the
primary OB star in each of the mentioned systems is primordial.
Any binary interaction, including tides, would not result in high
rotation velocity of the primary star in such short-period systems
(see \Secref{sec:pre-interaction}).

The formation of such rare binary configurations could be explained by
protostellar disk fragmentation and further inward migration while
accreting material at very early ($<$ 0.1 Myrs) evolutionary stages as was suggested by \cite{Tokovinin_2020}. However, we do not have any
observational constrain regarding the resulting distribution of stellar spin in
this scenario. Such a distribution has only been studied for the
pre-MS stars in the low-mass domain \citep[see
e.g.,][]{Serna_2021,Kounkel_2023}, where longer lifetimes increase
the importance of tides.

For the sake of completeness, we also checked the Multiple Star Catalog \citep{Tokovinin_2018} in order to find any short period systems with a mass of primary components $>$ 7 M$_{\odot}$ and $q$ $<$ 0.2.
The result of this query consisted of 19 binary systems that are either detached binary systems or hierarchical triples or quadruples, none of these systems having stars with \vsini~ > 200 \kms.
There are also short-period overcontact double-line spectroscopic
binaries \citep[see, e.g.][]{Abdul_Masih_2022}, which in most cases
appear to be tidally locked. These may appear as fast rotators as
well \citep[for example MY Cam and CC Cas, see][respectively]{Lorenzo_2014,Southworth_2022}.

We should also mention a few known short-period post-interacting OB binaries. The systematized properties of selected well-studied post-interacting Be/Bn binaries can be found in \citet{Harmanec_2015,Naze_2022_be,Schurman_2022, Wang_2023_stripped}.
Only a few systems with a period less than 10 days are known, and they are all rotating slower than our threshold: MX Pup \citep[\vsini~$\sim$ 120 \kms, $P$ $\sim$ 5.15 days,][]{Carrier_2002}, HD\,15124 \citep[\vsini~$\sim$ 95 \kms, $P$ $\sim$ 5.47 days, companion is a sdOB progenitor,][]{Badry_2022} and NGC1850 BH1 \citep[$P$ $\sim$ 5.04 days, which leads to \vsini~ $\sim$ 50 -- 70 \kms~ by assuming the period of ellipsoidal variability and $R$ $\sim$ 5--7 $R_{\odot}$, see][]{Badry_BH1,Saracino_2023}.
There also exist semi-detached B-type binaries with such short periods \citep[see Table 7 in][]{Harmanec_2015} with an ongoing mass transfer.
Nevertheless, none of the mentioned B-type systems has primaries rotating as fast as in the systems we focused on.

When considering longer period binaries, we note that there are more confirmed post-interacting companions in such short-period binaries with fast rotators of B-type than O-type.
Partly, this is caused by the observational difficulties to
detect a faint low-mass secondary component near a luminous massive O-type primary
star. For example, the majority of sdOB objects were detected through UV
spectroscopic observations \citep{Wang_2021}, where the flux ratio of
a hot stripped star with respect to the flux of a B-type star is
more favorable than for O-type binary systems \citep{Gotberg_2018}.

In summary, known OB short-period binaries could either have a pre- or
post-interacting companions, irrespective of the current rotation of
their primaries, but extreme mass ratio and fast rotation seem to be found only in the pre-interacting systems.
Thus, the fast-rotating tail observed for OB stars contains
some fraction of pre-interacting objects, suggesting that physical
  processes active during the pre-MS may contribute to the high tail of the
bimodal distribution of \vsini~\citep{vfts_2015_otype}.

In favor of this hypothesis, we can also point out that the rotation distributions
(up to \vsini~ $\sim$ 300 \kms) of apparently single O-type stars
and of primaries in the spectroscopic binaries have the same morphologies
\citep{vfts_2015_otype}. Considering our results on
the fastest rotator in the sample (HD\,46485), we can emphasize that
even in the regime of highest rotation, the pre-interacting systems can be found. Thus, fast rotation alone is not sufficient
  to confidently advocate for a mass-transfer origin of fast rotators and other observables, such as
chemical composition, apparent age, and kinematics \citep{blaauw:61, blaauw:93, Renzo_2019runaways, Renzo_2021}, should be
considered to clean samples of binary accretors and reveal the initial rotation distribution of massive stars.

\subsection{Theoretical implications}

Our interpretation of HD\,25631, HD\,191495, and HD\,46485 as
  short-period and extreme mass-ratio, massive binaries hosting a
  fast rotating primary since birth has several theoretical
  implications regarding star formation processes and early orbital
  evolution. As discussed above, rotation alone is not a
  sufficient smoking gun to establish the nature of an observed star
  as accretor, and a combination of signatures is desirable. These
  could include enhanced $^{14}\mathrm{N}$ and $^4\mathrm{He}$ at the
  surface \citep[e.g.,][]{blaauw:93, Renzo_2021}, ``rejuvenated'' age
  \citep[e.g.,][]{neo:77, schneider:16, renzo:23}, and possibly
  kinematic signatures \citep[e.g.,][]{Renzo_2019runaways}.

Also, our models suggest that formation processes can lead to primordial
  fast rotation, faster than initially tidally synchronized
  configuration where the orbital periods and rotational periods of
  both stars are equal,
  $P_\mathrm{orb}\equiv P_\mathrm{1, rot} \equiv P_\mathrm{2, rot}$.
  Observations show instead $P_\mathrm{1, rot}<P_\mathrm{orb}$, which
  implies an excess angular momentum in the stellar spin, and thus
  makes tides widen the orbit in the initial phase of evolution
  (\Secref{sec:pre-interaction}). Because the order of magnitude of spin angular momentum of a star
  (even at critical rotation) is much smaller than the orbital angular
  momentum, this is a relatively small effect, with typical orbital
  variations of $\Delta a\simeq$ 12\% \footnote{$\Delta$ $a$ calculated as the maximum relative variation in orbital separation that we found among our systems (namely, in the case of HD\,25631) at the moment the primary started the RLOF stage, when comparing the use of a $v_\mathrm{rot}^\mathrm{1,ini}$ = 1 \kms~ to that of $v_\mathrm{rot}^\mathrm{1,ini}$ = 220 \kms.}. This
  can cause a minor shift of boundary between case A and case B RLOF
  depending on the initial spin of the primary (see
  \Figref{fig:v_surf_hr}), potentially adding two extra dimensions
  (the initial rotation rate of both stars) to the parameter space for
  the evolution of binary systems.
  Moreover, this small increase in orbital separation $a$, although
  likely negligible, goes into the opposite direction compared to the
  ``unspecified'' orbital shrinking mechanism postulated to explain
  the early evolution of the radial velocity dispersion across young clusters
  \citep[e.g.,][]{sana:17, ramirez-tannus:21}.

  Tidal interactions in close binaries are a longstanding open
  question in stellar hydrodynamics. With the algorithms tested here
  (radiatively damped tides following \cite{Hut_1980} and
  layer-by-layer damping following \cite{Qin_2018}), and the low
  mass-ratios $q\lesssim0.15$ observed, tides are not able to
  significantly affect the spin of the primary star within the young
  inferred age. These algorithms have been compared before
  \citep[e.g.,][]{Qin_2018} and may not be representative of the full
  range of theoretical uncertainties \citep[e.g.,][]{Ma_2023, Sciarini_2024},
  however, in both cases tides seem to be insufficient to
  significantly affect the observed rotation rate within the lifetime
  estimated for the primaries observed. One key uncertainty is that
  models of tidal synchronization are based on ZAMS models showing no
  core-envelope boundary: including the evolution of the stellar
  structure is expected to weaken the effect of tides further,
  strengthening our conclusions \citep{Mirouh_2023}.
  Another factor to investigate how dynamical and equilibrium tides are affecting the synchronization and circularization timescales at different metallicity while the system is evolving  \citep[see,][]{Siess_2013}.

\section{Conclusions}
\label{sec:conclusions}
In the present work, we evaluate different evolutionary scenarios
that could explain the observed properties of three short-period
binaries investigated by \citet{Naze_2023_rot}. Specifically, we
  consider three options for the origin of the
  faster than tidally synchronized rotation rate, spectroscopically
  observed in the SB1 systems HD\,25631, HD\,191495, and HD\,46485: (\emph{i})
  star formation processes (the primordial and pre-interaction scenario), (\emph{ii})
  the result of spin-up due to mass transfer (post-interaction
  scenario), or (\emph{iii}) the result of a merger in an originally triple
  system. The orbital architecture of the binaries (short observed period $P<10$\,days,
  small-to-null eccentricity $e\simeq0$, and very low mass ratio $q<0.20$) put
  strong constraints on the possible origin of the observed fast rotation.

Our analysis shows that:

\begin{itemize}
  \item Primordial fast rotation can persist during most of the MS stage, as the effect of the tidal interaction only starts to be significant close to the terminal age when the radius of the primary component is increasing.
  \item Initial rotation affects the orbital configuration through the tidal interaction, widening the orbit in case of a large initial rotation thereby affecting when the mass transfer starts (before or after the MS stage, i.e., case A or case B mass transfer).
  \item The evaluation of the post-interacting scenario leads us to the conclusion that during conservative mass transfer, the orbit widens significantly so this type of interaction cannot reproduce the present observational properties of the investigated systems (especially in terms of $P$ and $q$).
  \item Various regimes of nonconservative mass transfer can lead to
    short-period systems; however, it is unlikely to get an extreme mass ratio configuration before the systems merged as the orbit is shrinking fast.
  \item An initially stable triple system may possibly result in a
    fast-rotating primary star after the merging of the inner binary.
    However, our current theoretical understanding suggests
      that MS mergers may be slow rotators and, most
      importantly, $\sim95\%$ of the simulated triple systems have a probability of being
      dynamically unstable $\gtrsim 90\%$. This suggests that within
      $\lesssim$ 100 outer periods ($\lesssim 2\,\mathrm{years}$ for
      our systems), the triple system would dynamically disrupt: this leaves
      an extremely narrow window for the merger to happen, and this would
      effectively reduce scenario (\emph{iii}) to scenario (\emph{i}) with the merger
      being part of the star formation process. 
\end{itemize}

Thus, we suggest that it is likely that the primaries in our three
systems have a fast-rotating primary from birth, that is, scenario (\emph{i})
is preferred. These results support the protostellar disk fragmentation mechanism of forming short-period binaries suggested by \citet{Tokovinin_2020}.
The detection of such young pre-interacting binaries can shed light on the overall initial spin distribution of massive stars.
In addition, if fast rotation can be primordial, it naturally implies that massive
stars could have a wide range of initial rotations, and that rotation
alone is not a sufficient indicator to uniquely identify past binary interactions.

\begin{acknowledgements}
We thank the anonymous referee for helpful comments that
have improved the manuscript. The authors acknowledge M.~Zapartas for the help with the tidal synchronization prescription that is implemented in the POSYDON code.
N.B. acknowledges support from the postdoctoral program (IPD-STEMA) of
Liege University, and the Belgian federal government grant for Ukrainian
postdoctoral researchers (contract UF/2022/10). Y.N. and G.R acknowledge support from the Fonds
National de la Recherche Scientifique (Belgium), the European Space
Agency (ESA), and the Belgian Federal Science Policy Office (BELSPO)
in the framework of the PRODEX Programme (contract linked to
XMM-Newton). M.R. thanks M.~Cantiello and A.~Trani for helpful discussion.
Computational resources have been provided by the Consortium des Équipements de Calcul Intensif (CÉCI), funded by the Fonds de la Recherche Scientifique de Belgique (F.R.S.-FNRS) under Grant No. 2.5020.11 and by the Walloon Region.
\end{acknowledgements}

\bibliographystyle{aa}
\bibliography{ref}

\begin{appendix}

\section{MESA setup}
\label{sec:MESA_setup}

We use MESA version 15140 to compute our models (MESA SDK version: x86$\_$64-linux-22.6.1). For all the stellar (and binary) models presented in
this study we assume the following physical and numerical parameters
unless otherwise specified. The MESA equation of
state (EOS) is a blend of the OPAL \citet{Rogers2002}, SCVH
\citet{Saumon1995}, PTEH \citet{Pols1995}, HELM \citet{Timmes2000},
and PC \citet{Potekhin2010} EOSes.

OPAL \citep{Iglesias1993, Iglesias1996} provides the main radiative
opacities, with low-temperature data from \citet{Ferguson2005} and the
high-temperature from \citet{Buchler1976}. Electron conduction
opacities are from \citet{Cassisi2007}.

Nuclear reaction rates are a combination of rates from NACRE
\citep{Angulo1999}, JINA REACLIB \citep{Cyburt2010}, plus additional
tabulated weak reaction rates \citet{Fuller1985, Oda1994,
  Langanke2000}. Screening is included via the prescription of
\citet{Chugunov2007}.  Thermal neutrino loss rates are from
\citet{Itoh1996}. We use a
22-isotope nuclear network (\texttt{approx\_21\_plus\_cr56}).

We used a mixing length parameter of 1.5 and convective stability defined by
\citet{Ledoux_1947} criterion to treat convective mixing. Overshooting
is included for any core convection phase, using an exponentially
decay functional form \cite{Herwig_2000} with free parameters
suggested by \cite{claret:18}. We also assume a constant rotational frequency along
isobaric surface \citep{Ekstrom2012}. In addition, rotation
implementation in MESA includes the Goldreich-Schubert-Fricke (GSF)
instability \citep{Heger_2000} and Eddington–Sweet circulations
\citep{Sweet_1950}. Following \citet{Heger_2000} we assume the
coefficient of rotational mixing (\texttt{am\_D\_mix\_factor}) that
took into account several rotational components equal to 0.33.
By default for all models, we assume a Spruit - Tayler dynamo (\texttt{am\_nu\_ST\_factor = 1.0}) for the transport of angular momentum \citep{spruit:02}.
The inlists, processing scripts, and model output are available at~\url{https://zenodo.org/records/10479754}.

\section{Resolution tests}
\label{sec:resolution_test}

\begin{figure}[!ht]
  \centering
  \script{plot_save_mesa_comparison_wind_fin.py}

  \includegraphics[width=0.45\textwidth]{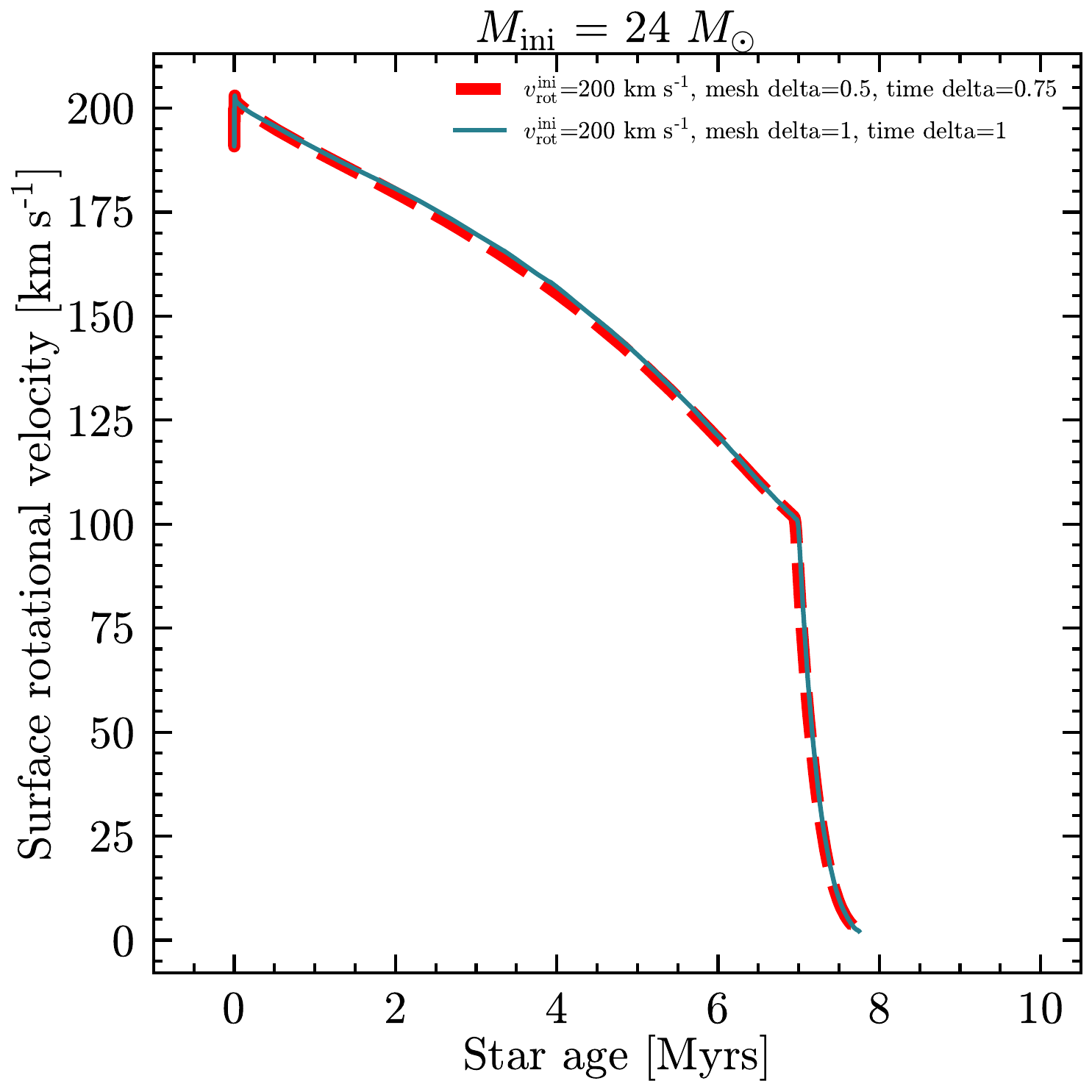}
  \includegraphics[width=0.45\textwidth]{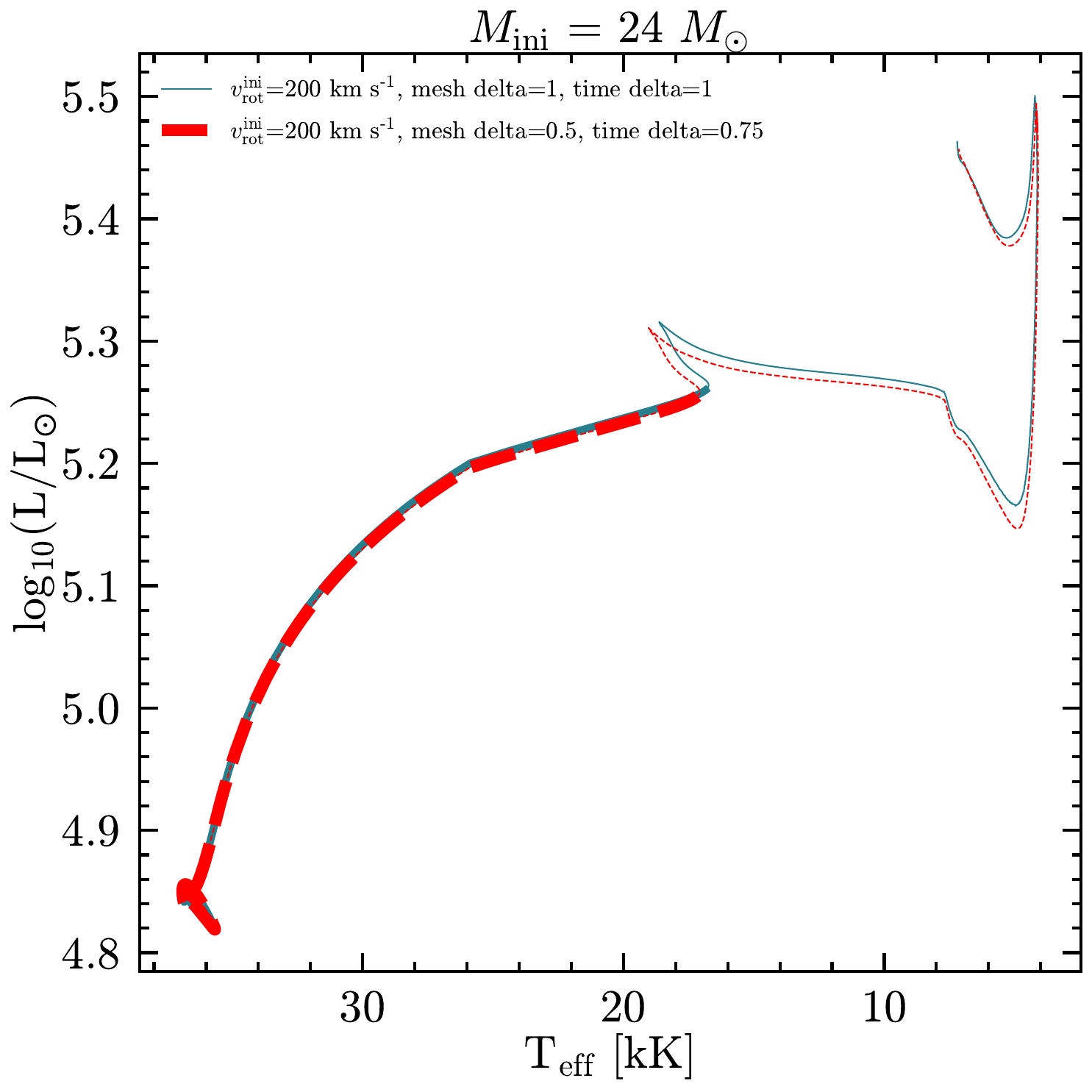}
  \caption{Evolution of average equatorial rotation
    velocity for a single 24 $M_{\odot}$ star with $v_\mathrm{rot}^\mathrm{ini}$ =
    200 \kms~ adopting different \texttt{mesh\_delta\_coeff} and
    \texttt{time\_delta\_coeff} parameters ({\bf top panel}). The tracks continues only
    up to the terminal-age evolutionary phase.  {\bf Bottom panel:}
     HR diagram for the same model that continues until the end of
    the He-core burning phase.}
  \label{fig:resolution_test_one_star}
\end{figure}

\begin{figure}[!ht]
  \centering
  \script{plot_save_mesa_comparison_panels_fin.py}
  \includegraphics[width=0.45\textwidth]{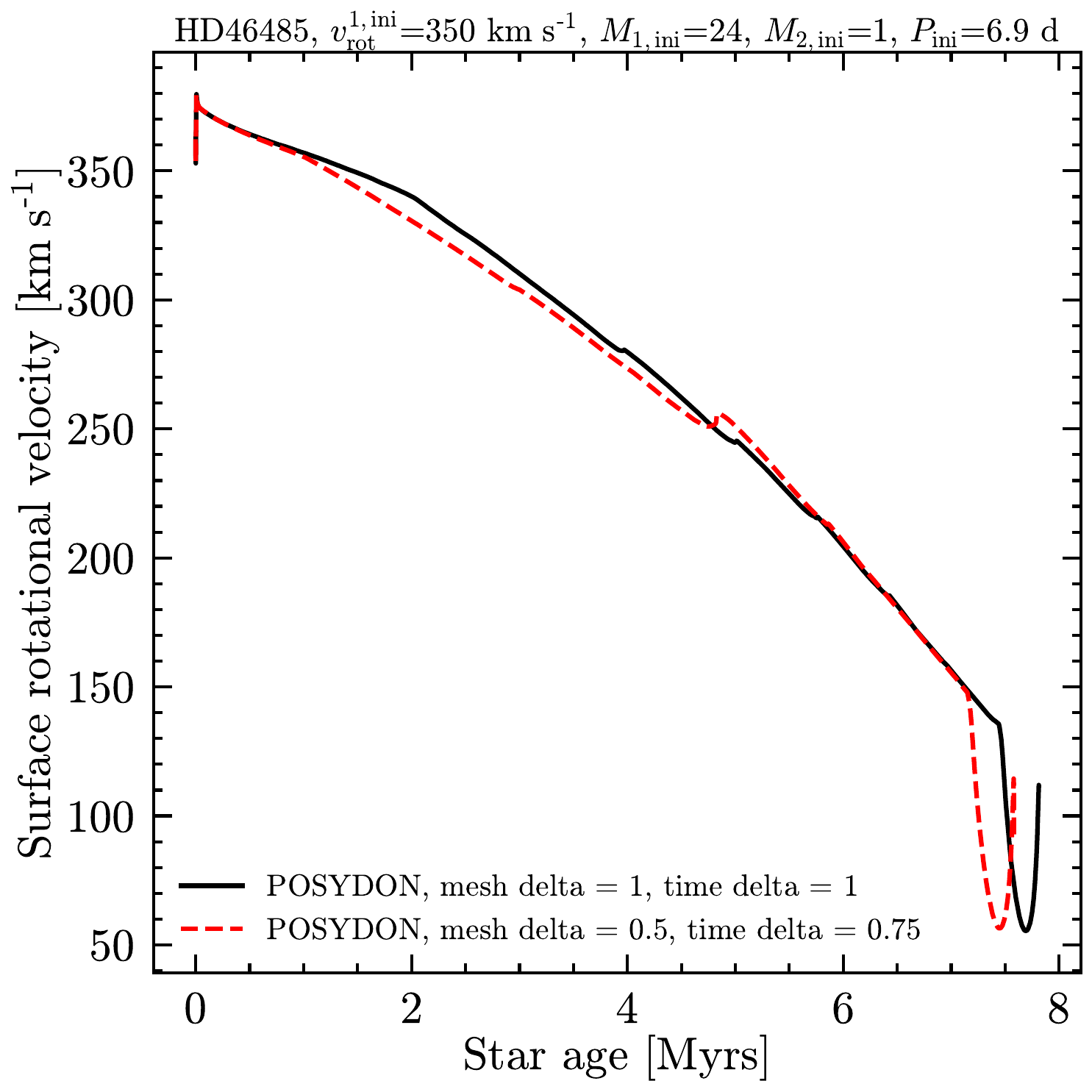}
\includegraphics[width=0.45\textwidth]{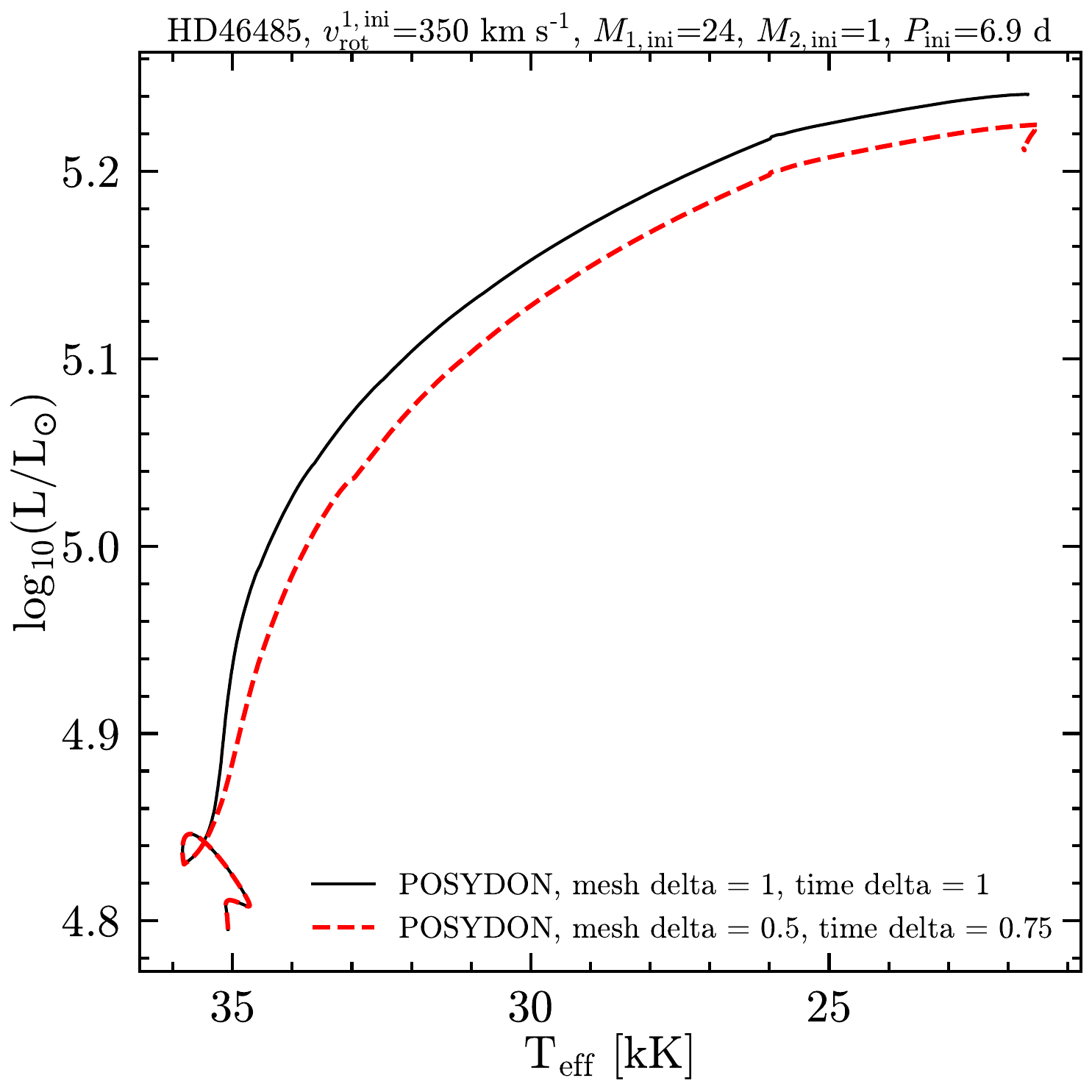}
  \caption{Same as \Figref{fig:resolution_test_one_star}, but for the HD\,46485 system by taking into account the POSYDON structure dependent tidal synchronization prescription.}
  \label{fig:resolution_test_tides}
\end{figure}

\begin{figure}[!ht]
  \centering
  \script{plot_save_mesa_individual_pos_res.py}
  \includegraphics[width=0.45\textwidth]{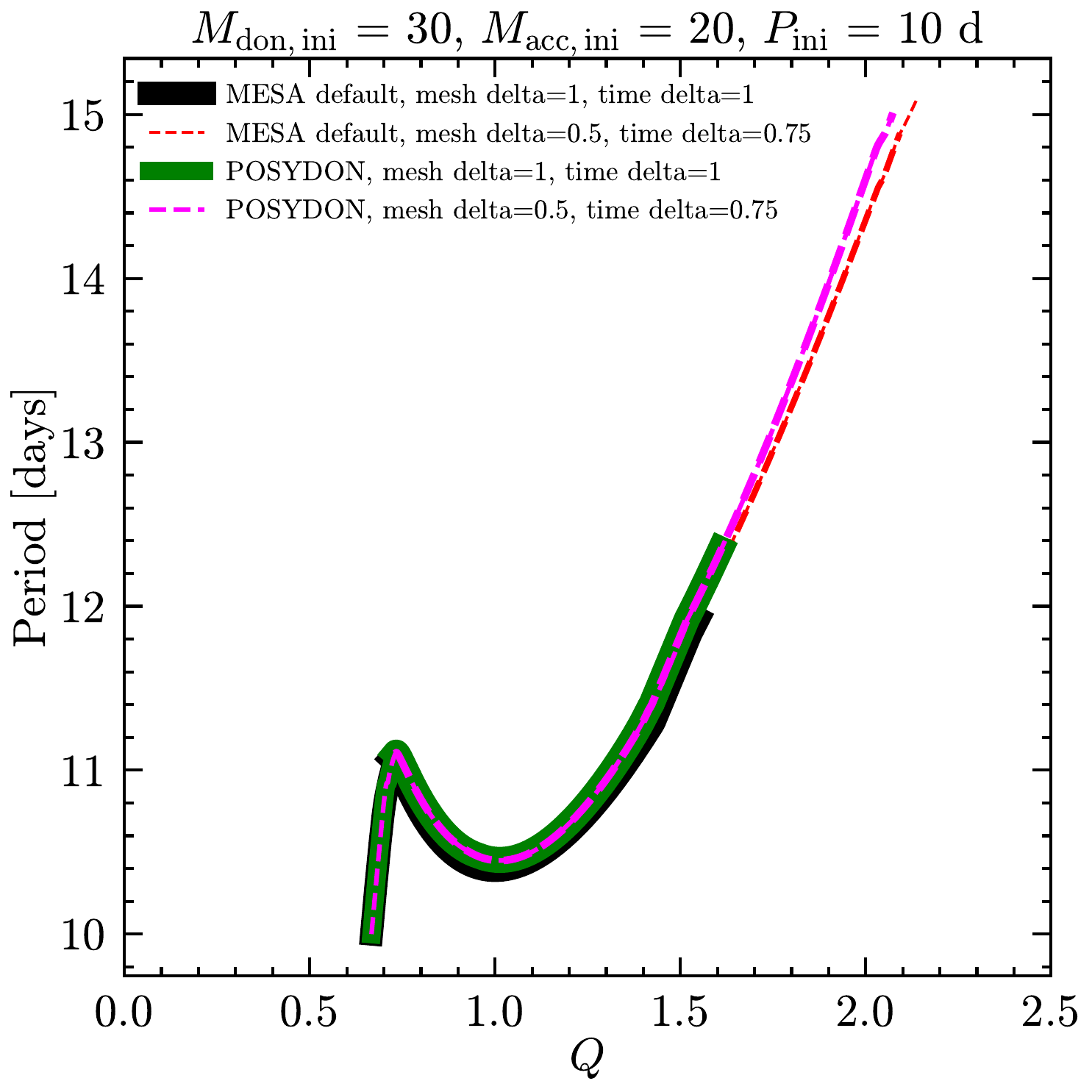}
  \caption{Period as a function of the mass ratio for a system initially defined as $M_\mathrm{don,ini}$ = 30 $M_{\odot}$, $M_\mathrm{acc,ini}$ = 20 $M_{\odot}$, and $P_\mathrm{ini}$ = 10 d. The distribution includes the MESA default and POSYDON tidal synchronization prescriptions.}
  \label{fig:resolution_test_pvsq}
\end{figure}

\begin{figure}[!ht]
  \centering
  \script{plot_save_mesa_pos_eta_res.py}
  \includegraphics[width=0.49\textwidth]{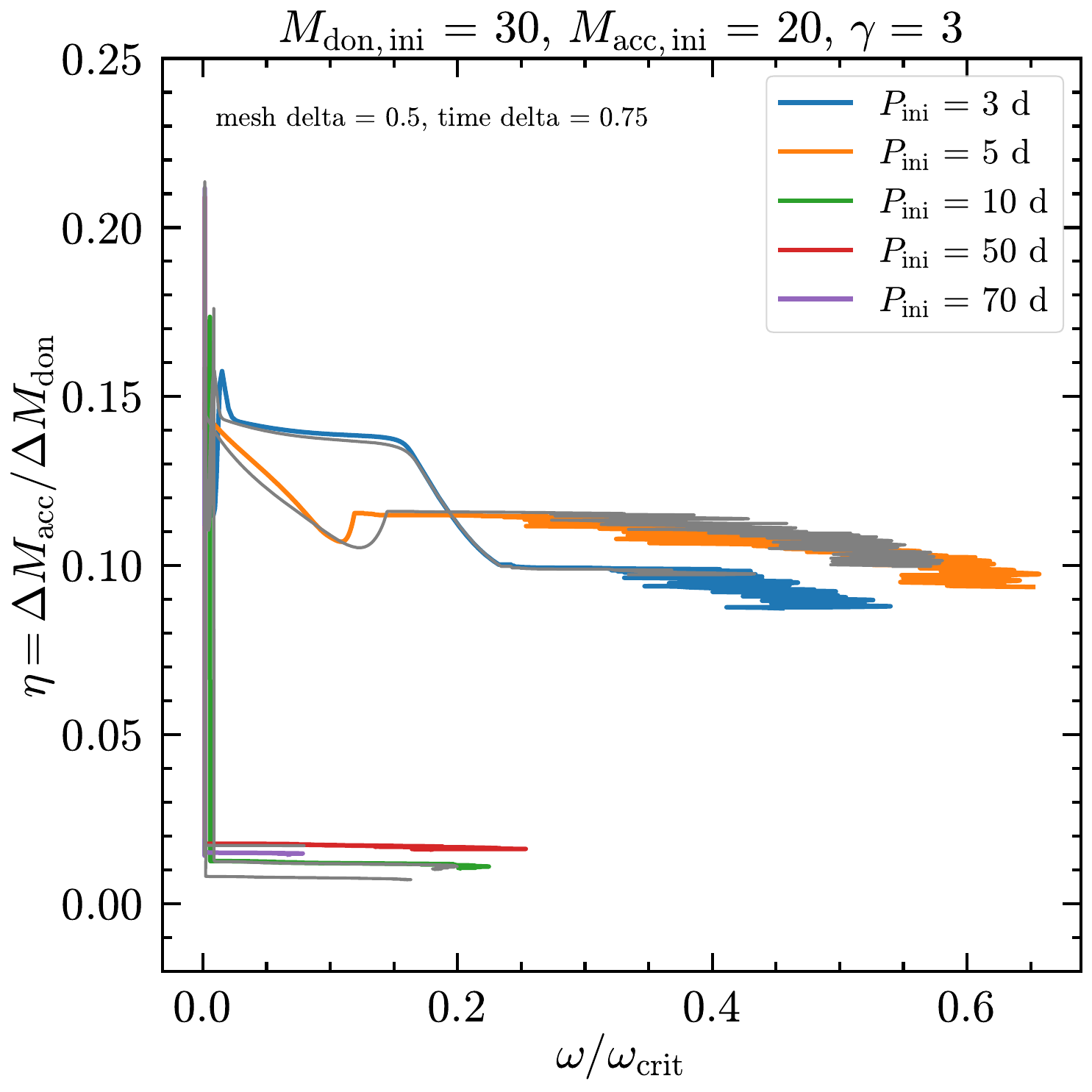}
  \caption{Mass transfer efficiency as a function of the angular velocity for the various initial periods of a given system (by taking into account the POSYDON tidal synchronization prescription). The gray points represent the model with the \texttt{mesh\_delta\_coeff} and \texttt{time\_delta\_coeff} equal to 1 and the "MESA default" tidal synchronization prescription.}
  \label{fig:eta_omega_res}
\end{figure}

We performed a resolution test (numerical convergence) for each of the single star and binary interaction scenarios. To do so, we increased the spatial and time resolution by decreasing the \texttt{mesh\_delta\_coeff} and \texttt{time\_delta\_coeff} parameters respectively.
The \texttt{mesh\_delta\_coeff}  represents the maximum of allowed deltas for each grid point, in our tests we decreased it by a factor of two from 1 to 0.5. In addition, we decreased by 25\% the \texttt{time\_delta\_coeff} that allowed us to make a more dense grid in terms of the time interval between the models.
Namely, we increased the average number of zones for each model (mesh points) from $\sim$1600 to $\sim$3100, and decreased by one-third the average time step per model (i.e., from $\sim$3200 to $\sim$2300 years).
The results of these tests are presented in Figures \ref{fig:resolution_test_one_star}, \ref{fig:resolution_test_tides}, \ref{fig:resolution_test_pvsq}, and \ref{fig:eta_omega_res}. As can be seen in these figures, the change is smaller than observational uncertainties, highlighting the reliability of our results.

\section{Different formalisms of defining the tidal $E_{2}$ coefficient.}
\label{sec:e2_def}
\begin{figure}[!ht]
  \centering
  \includegraphics[width=0.50\textwidth]{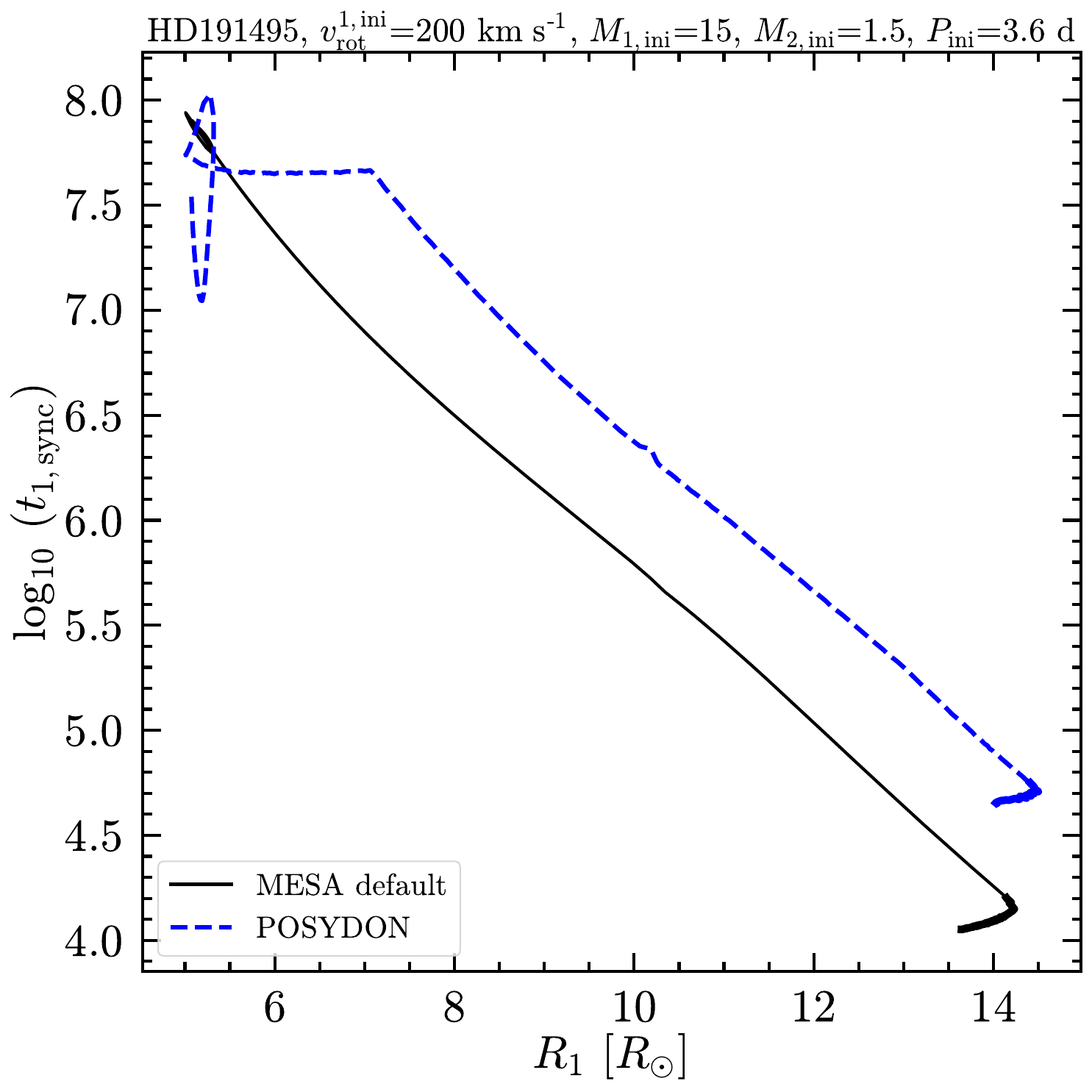}
  \caption{Synchronization timescale as a function of the primary component radius of HD\,191495 system by assuming the MESA default (\texttt{sync\_type = "Hut$\_$rad"}) and POSYDON (\texttt{sync\_type = "structure\_dependent"}) tidal prescriptions.}
  \label{fig:lg_tsync}
  \script{plot_save_mesa_comparison_panels_fin.py}
\end{figure}

To emphasize how the strengths of tides are sensitive to the adopted underlying physics we calculated the synchronization timescales ($t_\mathrm{sync}$) for one of our stars where such differences are the most prominent because of the short orbital period (i.e. HD\,191495, $P \sim$ 3.6 d).
We illustrate in \Figref{fig:lg_tsync} how the synchronization timescale varies within the entire duration of the MS as a function of the radius of the primary component.
The observed significant difference in $t_\mathrm{sync}$ for a given radius of the star (that also represents the evolutionary stage on the MS) occurs because of the different ways of calculating the $E_{2}$ coefficient which is important for calculating the timescale of synchronization for dynamical tides (regime of tides in the stellar radiative envelopes, $(1/t_\mathrm{sync, radiative} \sim E_{2}$).
Indeed, as it addopted in \citet[]["MESA default" prescriprions]{Hurley_2002} $E_\mathrm{2} = 1.592 \times 10^{-9} (M/M_\mathrm{\odot})^{2.84}$, while in \citet[]["POSYDON" formalims for H-rich stars]{Qin_2018}: $E_\mathrm{2} = 10^{-0.42} (R_\mathrm{conv}/R)^{7.5}$.
As we can see, in the "MESA default" prescription the $E_\mathrm{2}$ parameter depends only on the entire mass of a star, while in the POSYDON formalism it depends on the stellar radius and the radius of the convective core ($R_\mathrm{conv}$).
Thus, the differences in resulting synchronization timescales which are presented in \Figref{fig:lg_tsync} occur because in the POSYDON prescription the $E_\mathrm{2}$ is very sensitive to the size of the convective zone, especially taking into account that the ratio of $R_\mathrm{conv}/R$ is in a power of 7.5.
In this test, we are illustrating how tidal interactions are sensitive to the adopted physics of tides and the evolutionary stage of the stars especially in massive binaries with an extended radiative envelope.
Thus, while modeling the evolution of massive binaries, we suggest using the POSYDON tides formalism based on \citet{Qin_2018} study.

\section{Synchronization and circularization of the orbits}
\label{sec:angular_vel}

\begin{figure}[!ht]
  \centering
  \includegraphics[width=0.50\textwidth]{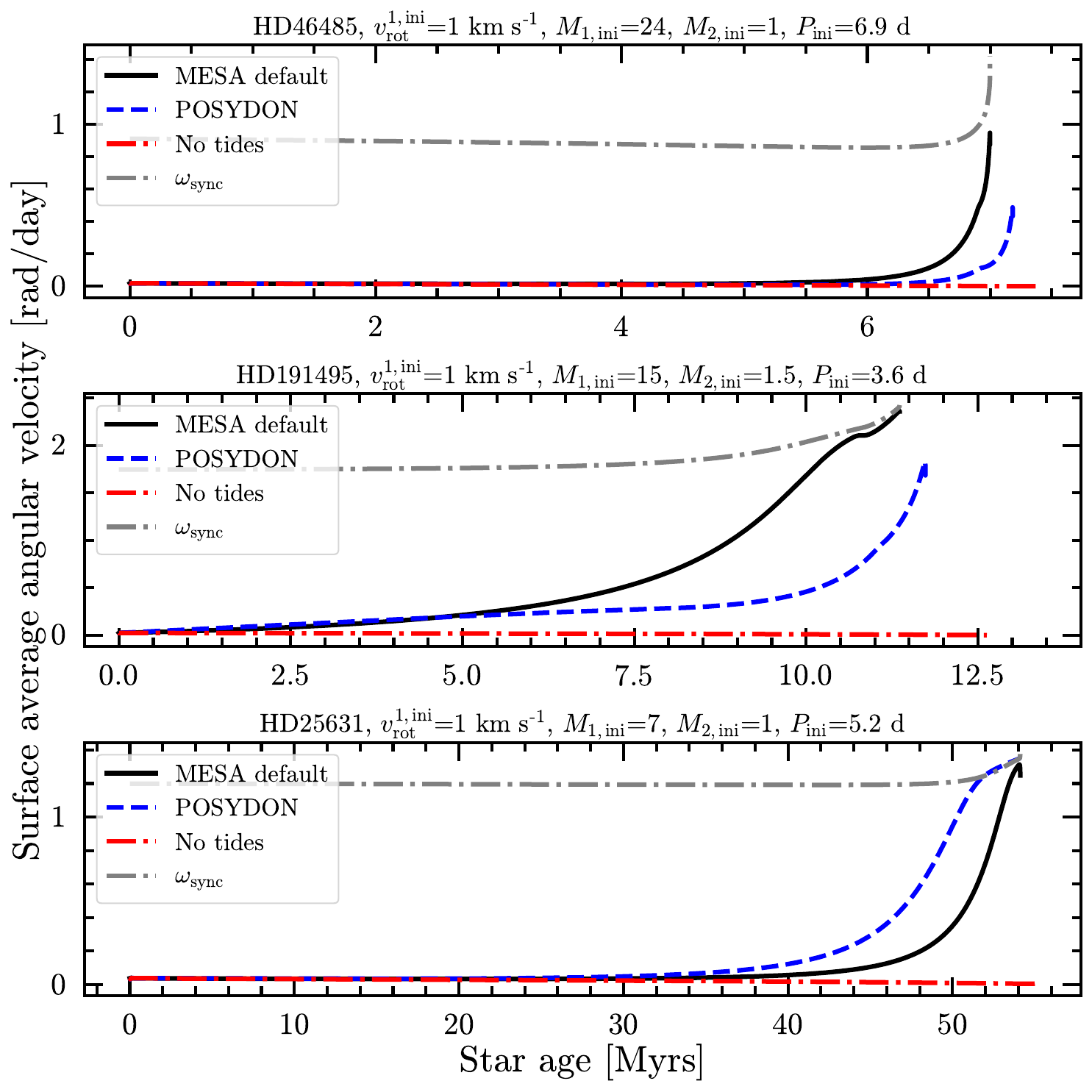}
  \caption{Evolution of surface average angular rotational velocity for our
    three program stars by taking into account different tidal
    synchronization prescriptions (assuming $v_\mathrm{rot}^\mathrm{1, ini}$ = 1 \kms). "MESA default" refers to the
    radiative envelope tidal synchronization prescription (\texttt{sync\_type = "Hut\_rad"}), and "POSYDON" to the structure dependent
    prescription.}
  \label{fig:v_omega}
  \script{plot_save_mesa_comparison_panels_acc_fin.py}
\end{figure}

In \Figref{fig:v_omega} we present the evolution of surface average angular velocity for the three studied systems when assuming $v_\mathrm{rot}^\mathrm{ini}$ = 1 \kms.
We illustrate its behavior by using different tides prescription, namely \texttt{sync\_type = "Hut\_rad"} \citep{Hut_1980,Hurley_2002}, \texttt{sync\_type = "structure\_dependent"} \citep{Qin_2018}, and no tidal interactions at all.
The synchronization angular velocity is calculated as $\omega_{sync} = 2 \pi / P$. Where, $P$ is the orbital period of the system that represents the orbit variations by assuming \texttt{sync\_type = "Hut\_rad"}.

In order to investigate the typical circularization timescale for our system, we used \texttt{circ\_type = "Hut\_rad"} for both components which is based on the circularization timescales for radiative envelopes following \citet{Hurley_2002}.
We assumed an initial eccentricity equal to 0.5.

\begin{figure}[!ht]
  \centering
  \script{plot_e_fin.py}
  \includegraphics[width=0.50\textwidth]{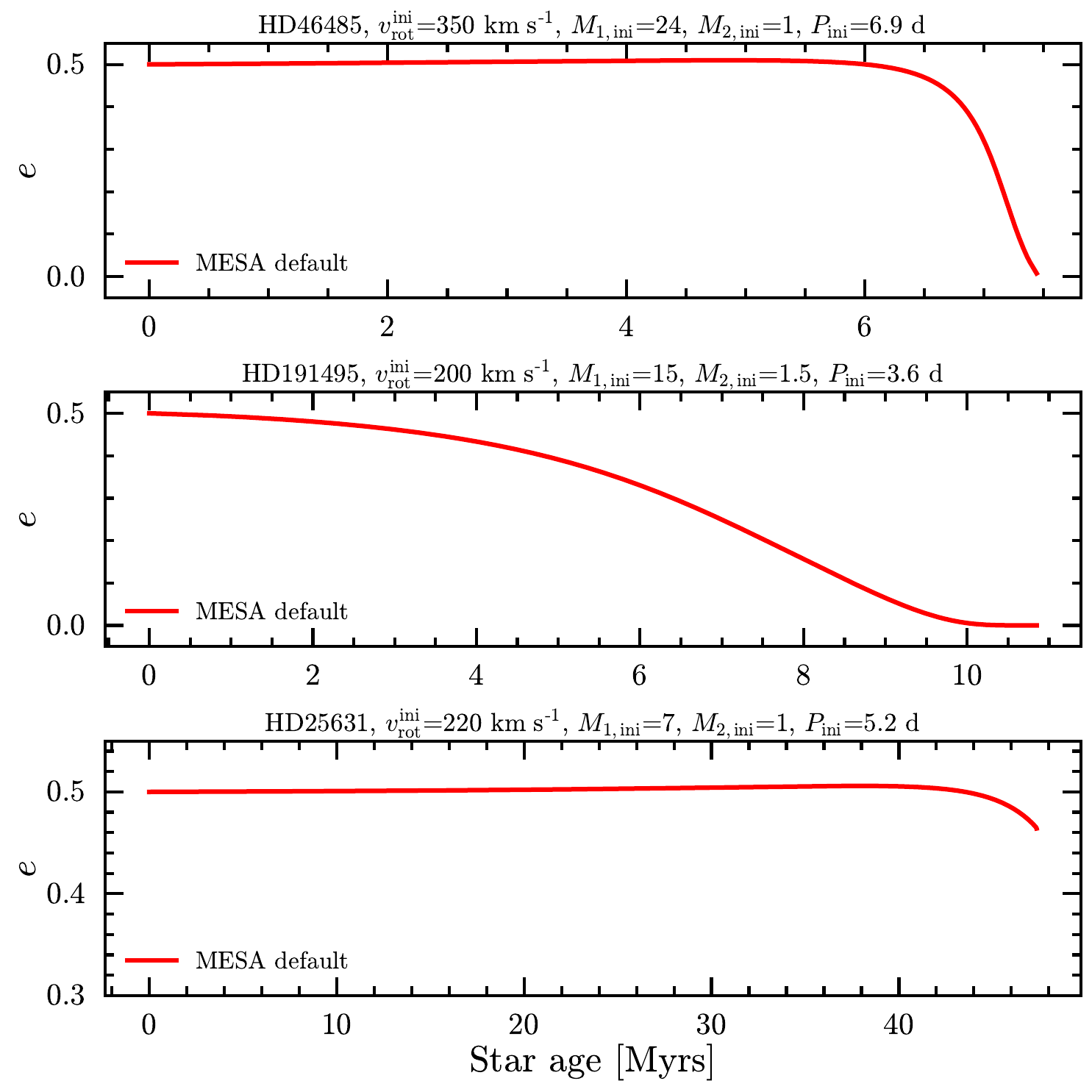}
  \caption{Evolution of eccentricity for our
    three stars assuming initial eccentricity equal to 0.5. "MESA default" refers to the
    \texttt{sync\_type = "Hut\_rad"} tidal synchronization prescription.}
  \label{fig:v_ecc}
\end{figure}

\end{appendix}

\end{document}